\documentclass[aps,prd,nofootinbib,twocolumn,superscriptaddress,preprintnumbers,balancelastpage,longbibliography,nobibnotes]{revtex4-1}

\usepackage{xcolor}
\usepackage{rotating} 
\usepackage{amsmath,amssymb,mathtools,bm}
\usepackage{graphicx, color, hepunits}
\usepackage[colorlinks=true
,urlcolor=blue
,anchorcolor=blue
,citecolor=blue
,filecolor=magenta
,linkcolor=red
,menucolor=blue
,linktocpage=true
,pdfproducer=medialab
,pdfa=true
]{hyperref}
\usepackage[utf8]{inputenc}
\usepackage[english]{babel}

\begin{document}

\title{Exotic Particles at the DUNE Near Detector from Charged Pion Scattering}

\author{Diana Forbes}
\affiliation{Department of Physics, University of Illinois Urbana-Champaign, Urbana, IL 61801, U.S.A.}
\affiliation{Illinois Center for Advanced Studies of the Universe, University of Illinois Urbana-Champaign, Urbana, IL 61801, U.S.A.}

\author{Yonatan Kahn}
\affiliation{Department of Physics, University of Illinois Urbana-Champaign, Urbana, IL 61801, U.S.A.}
\affiliation{Illinois Center for Advanced Studies of the Universe, University of Illinois Urbana-Champaign, Urbana, IL 61801, U.S.A.}

\author{Rachel Nguyen}
\affiliation{Department of Physics, University of Illinois Urbana-Champaign, Urbana, IL 61801, U.S.A.}
\affiliation{Illinois Center for Advanced Studies of the Universe, University of Illinois Urbana-Champaign, Urbana, IL 61801, U.S.A.}

\begin{abstract}
Fixed-target proton-beam experiments produce a multitude of charged pions that rescatter in the beam dump. These charged pion scattering events can be an additional irreducible source of exotic particles which couple to photons or hadrons. We analyze the sensitivity of the DUNE Near Detector complex to millicharged particles (MCPs) and heavy axion-like particles (ALPs) with low-energy couplings to gluons. Using the framework of chiral perturbation theory, we demonstrate regimes of parameter space where the charged pion production channel dominates over previously-considered production mechanisms for both MCPs and ALPs, thereby improving the sensitivity of DUNE to these new particles compared to previous studies.
\end{abstract}
\maketitle

\section{Introduction}

In the search for light, weakly-coupled extensions to the Standard Model, fixed-target experiments have proved to be an extremely useful approach (see Ref.~\cite{Gori:2022vri} and references therein). Trading energy for luminosity compared to colliding-beam experiments, these $\mathcal{O}(10-100) \ {\rm GeV}$ proton and electron beams can deliver Avogadro's number of beam particles on target, allowing us to search for a wide variety of low-mass particles beyond the Standard Model (BSM) with small production cross sections. In particular, proton beams produce copious amounts of secondary particles, including neutral mesons and muons, which can generate BSM particles through rare decays, mixing, or bremsstrahlung. Recent studies~\cite{Kelly:2018brz,Harnik:2019zee,Kelly:2020dda,Berger:2024xqk} have demonstrated that as part of the upcoming DUNE program~\cite{DUNE:2015lol,DUNE:2016evb,DUNE:2016hlj,DUNE:2016rla}, the neutral meson channel yields excellent sensitivity to millicharged particles (MCPs) (denoted $\chi$) or heavy axion-like particles (ALPs) coupled to QCD (denoted $a$).

However, previous analyses have overlooked an additional production mechanism for these new particles, namely charged pion scattering. At DUNE, the 120 GeV proton beam creates $\sim 6.5\,  \pi^{\pm}$ for every proton on target (POT) compared to $\sim 3.5 \, \pi^0$ per POT \cite{Berlin:2018pwi}. As a charged pion traverses the target, it can scatter off the target nuclei and produce BSM particles through bremsstrahlung-like processes, which may be described in the framework of chiral perturbation theory as long as the BSM particles have masses well below $4\pi f_\pi \sim 1.2 \ {\rm GeV}$, where $f_\pi = 93 \ {\rm MeV}$ is the pion decay constant. This production channel was first considered for dark photon production at DarkQuest \cite{Curtin:2023bcf}, where it was shown that charged pion bremsstrahlung generated comparable event rates to the Drell-Yan production channel for GeV-scale dark photons.

In this paper, we follow the logic of Ref.~\cite{Curtin:2023bcf} to search for MCPs and heavy ALPs at DUNE through secondary charged pion scattering. The MCP production channel is very similar to the previously-studied dark photon production, involving an off-shell virtual photon generating a $\chi \bar{\chi}$ pair instead of an on-shell dark photon decaying to visible states, and we will show that charged pion scattering does indeed yield modest gains in sensitivity for GeV-scale $\chi$. For heavy axions, QCD-coupled axion production from charged pions dominates the meson mixing production channel except in narrow resonance windows around the neutral meson masses, and is comparable to the expected rate from kaon decays~\cite{Berger:2024xqk}. At high axion masses, charged pion scattering provides additional sensitivity at large couplings all the way up to the regime of validity of chiral perturbation theory.

This paper is organized as follows. In Sec.~\ref{sec:detector}, we briefly describe the experimental design of the DUNE Near Detector (ND) Complex located at Fermilab, including additional detector components proposed for detecting MCPs. In Sec.~\ref{sec:simulation}, we give an overview of our simulation pipeline that we use to model the production and detection of BSM particles from charged pion scattering. In Sec.~\ref{sec:millicharged}, we describe our MCP model and calculate the additional sensitivity gained at DUNE ND at high MCP masses from charged-pion scattering, compared to the production modes previously considered in Refs.~\cite{Kelly:2018brz,Harnik:2019zee}. In Sec.~\ref{sec:axions}, we do the same for the heavy axion model of Ref.~\cite{Kelly:2020dda}. We conclude in Sec.~\ref{sec:conclusions}.

\section{DUNE as a BSM detector}
\label{sec:detector}

\begin{figure*}
    \centering
    \includegraphics[width=0.45\textwidth]{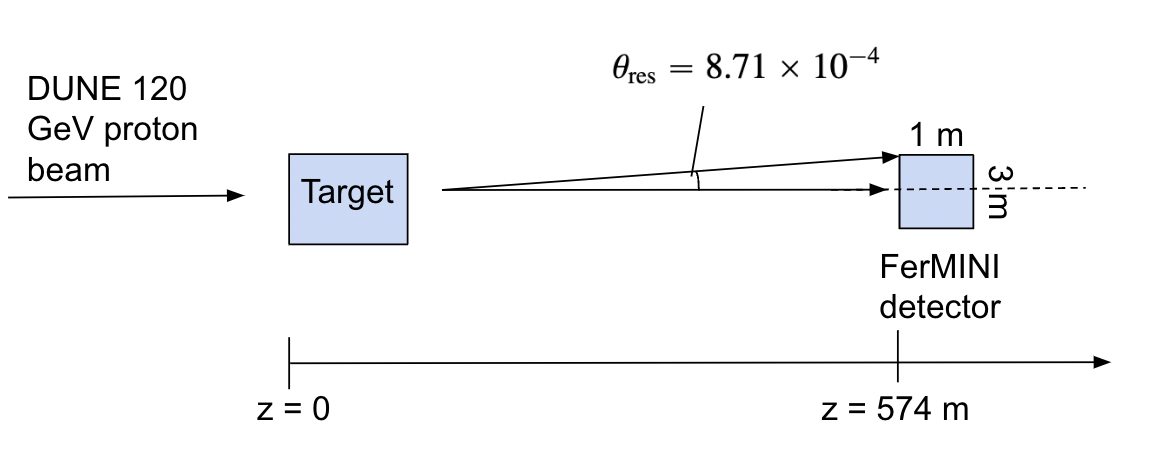}
    \includegraphics[width=0.53\textwidth]{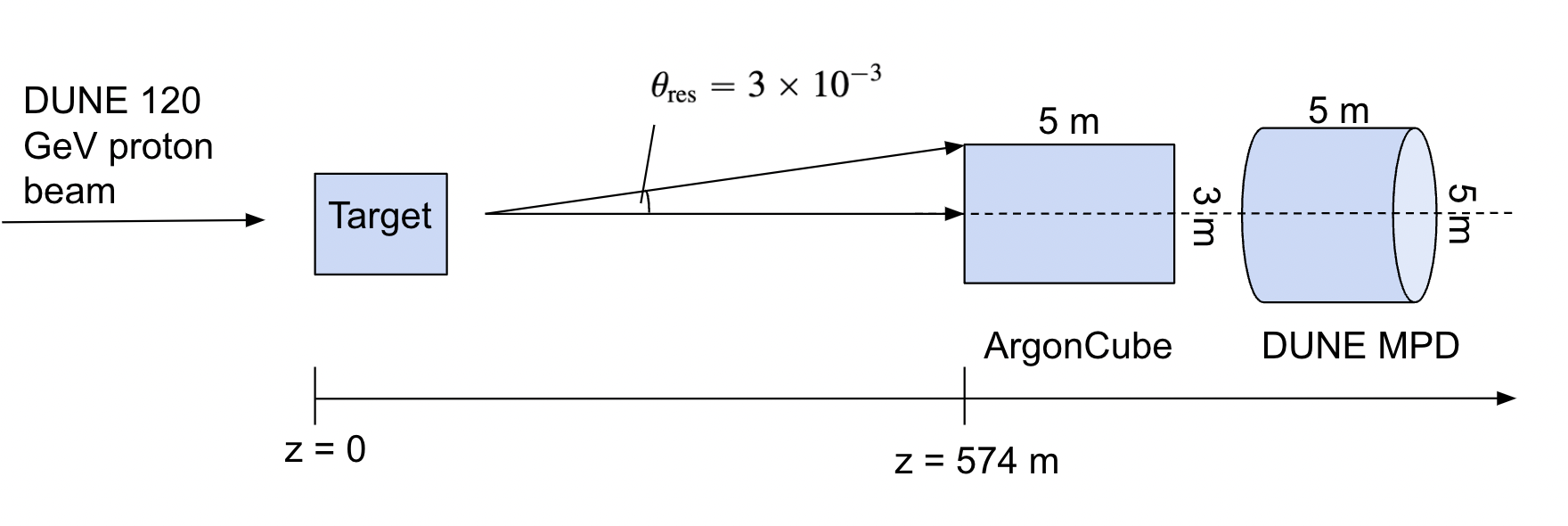}
   \caption{A simple schematic of the experimental setup for MCP (left) and ALP (right) searches. Note that the DUNE ND Complex contains ArgonCube and the Multi-Purpose Detector, which are the detectors used for our ALP search. Additionally, for the sake of our MCP search, one could imagine installing scintillator arrays and PMTs inside the Dune ND Complex inspired by the FerMINI proposal~\cite{Kelly:2018brz}, described in more detail in Sec.~\ref{sec:MCP_Signature}.
    \label{fig:DUNE ND Complex}
    }
\end{figure*}

DUNE is primarily designed to produce neutrinos, but its 120 GeV proton beam can also be used as a source for BSM particles, which can be detected at the DUNE Near Detector (ND) Complex. The LBNF-DUNE beam is a 120 GeV proton beam incident on a graphite target, with an expected luminosity of $1.47 \times 10^{21}$ protons on target (POT) each year, and a total of $1.47 \times 10^{22}$ POT over a 10-year run \cite{Berryman:2019dme,DUNE:2015lol,DUNE:2016evb,DUNE:2016hlj,DUNE:2016rla}. The ND Complex is located 574 meters down the beam line from the target, and will contain two argon detectors: ArgonCube, a 50 ton liquid argon time projection chamber (LArTPC), and the Multi-Purpose Detector (MPD), a gaseous argon time projection chamber designed to detect muons and other particles that are not stopped in ArgonCube. ArgonCube is 7 m in width, 3 m in height, and 5 m in length along the beamline. The MPD is situated directly downstream from ArgonCube and has a cylindrical volume that is 5 m in diameter and 5 m in height. Both ArgonCube and the MPD are sensitive to diphoton and hadron signatures by reconstructing the invariant mass and direction of the event, and these will be the primary decay modes of the heavy axions we consider.

To search for MCPs, one may also imagine adding scintillator arrays in the Near Detector Complex, as was proposed in~\cite{Kelly:2018brz}. To make a direct comparison with Ref.~\cite{Kelly:2018brz}, we consider the same setup: three stacks of scintillator arrays in a 1 m $\times$ 1 m detection area, coupled to photomultiplier tubes (PMTs) to detect charged particles. The distinctive signature of a MCP would be soft ionizations with yield below that expected from a minimum-ionizing particle of charge $|Q| = 1$.

In Fig.~\ref{fig:DUNE ND Complex}, we show a schematic of the DUNE ND complex and the detectors of interest. Relevant background processes for each of our BSM signatures will be discussed in Secs.~\ref{sec:millicharged} and~\ref{sec:axions}.

\section{Simulation Details}
\label{sec:simulation}

In order to calculate the production and detection of new particles at DUNE ND, we perform a Monte Carlo simulation. In this section, we outline the general simulation setup that generates BSM events from charged pion-nucleon scattering and estimates the detection efficiency within the detectors described above. In Secs.~\ref{sec:millicharged} and~\ref{sec:axions}, we will specify to MCPs and heavy axions, respectively.

\subsection{Charged Pion Production}
\label{subsec:chargedpionproduction}
 
When the 120 GeV DUNE proton beam collides with the target, it produces a multitude of SM mesons, including $\pi, \eta, \eta'$ as well as heavier mesons. While mesons are produced during several nuclear interaction lengths, in order to provide a conservative estimate of the reach and to make a direct comparison with previous studies~\cite{Berlin:2018pwi,Kelly:2018brz,Harnik:2019zee,Kelly:2020dda,Curtin:2023bcf}, we only consider the mesons produced within the first thin layer of the target. 

\begin{figure}
    \centering
    \includegraphics[width=0.48\textwidth]{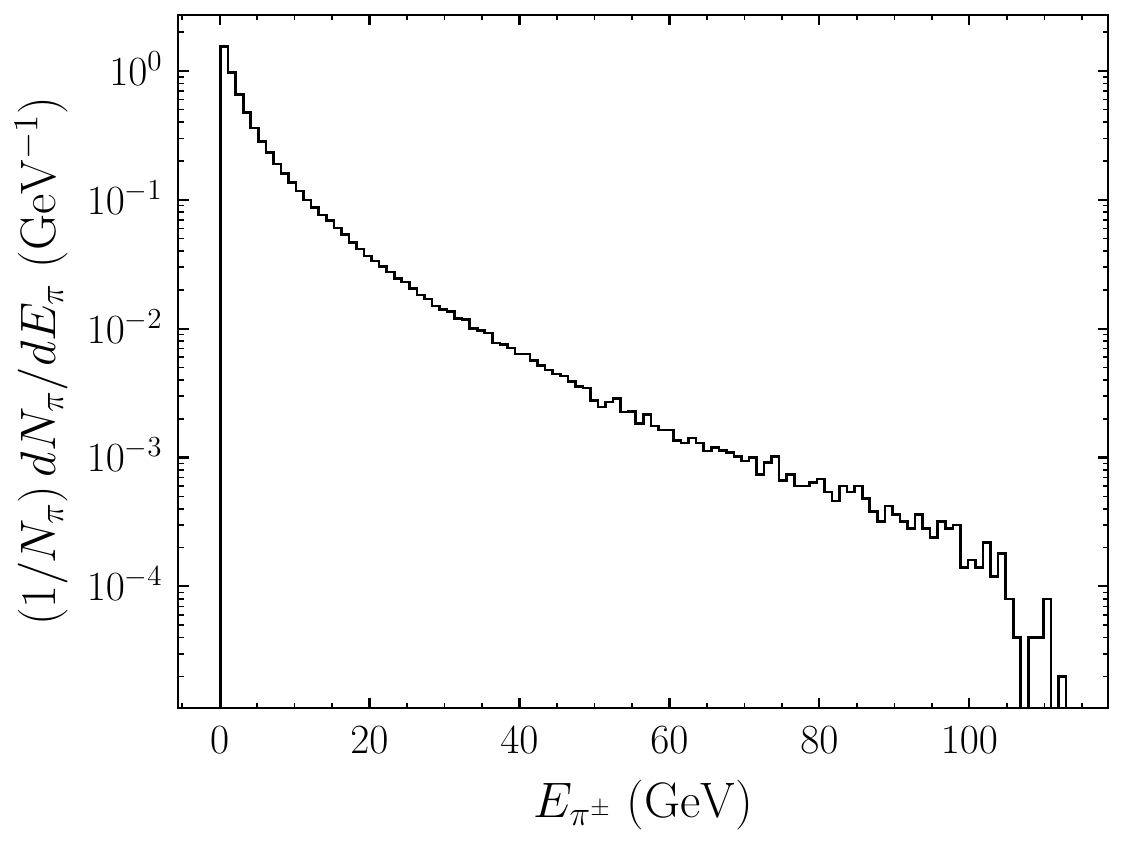}
    \caption{The energy distribution of charged pions from a \texttt{Pythia 8.2} simulation of a 120 GeV proton hitting a proton target. The pions were created within the first thin slice of the target. We rescale this distribution to match the nucleon content of the DUNE graphite target. This distribution was adapted from the analysis of \cite{Berlin:2018pwi}.}
    \label{fig:chargepiondist}
\end{figure}

To estimate the SM meson spectrum, we adapt the \texttt{Pythia 8.2} simulation of a 120 GeV proton incident on a proton target from \cite{Berlin:2018pwi}, rescaled to match the nucleon composition of the DUNE graphite target. The number of mesons produced per 120 GeV proton is $N_{\pi^0} \sim 3.5$, $N_{\pi^{\pm}} \sim 6.5$, $N_{\eta} \sim 0.40$, and  $N_{\eta'} \sim 0.04$.\footnote{Treating the target as free protons is clearly a very rough approximation, however it allows us to make a direct comparison with previous studies in the literature~\cite{Berlin:2018pwi,Kelly:2018brz}. Ref.~\cite{Kelly:2018brz} (also Ref.~\cite{Kelly:2020dda}) found $N_{\pi^0} \sim 4.5 \ (2.9)$, $N_\eta \sim 0.5 \ (0.33)$ for the DUNE target, which only differ by $\sim 20\%$ from our values. In the approximation where all meson production takes place in the first nuclear interaction length of the target, the target composition likely affects the production rates only weakly.} The total number of charged pions can be written as an integral over the energy spectrum of charged pions produced,
\begin{equation}
    N_\pi = \int dE_{\pi} \frac{d N_{\pi}}{d E_{\pi}}.
\end{equation}
For simplicity, we assume $N_{\pi^+} = N_{\pi^-} = N_{\pi^{\pm}}/2$.
The energy distribution of the charged pions is shown in Figure \ref{fig:chargepiondist}. The majority of charged pions are produced with low energy, $E_{\pi} \sim 1-10$ GeV. However, the high-energy tail of the distribution, combined with the large quantity of charged pions produced in the target (of order $10^{23}$ over the lifetime of DUNE) can produce a detectable number of BSM particles that are boosted enough to reach the DUNE ND.

\subsection{BSM Particle Flux}
\label{subsec:chargedpionproduction}

Exotic particles may be created from pion-nucleon scattering, $\pi^{\pm} N \to \pi^{\pm} N' X$, where we will later consider $X = a$ or $\chi \bar{\chi}$.\footnote{Axions can also be produced through pion conversion, $\pi N \to a N$, though we will discuss in Sec.~\ref{sec:axions} that this process is (surprisingly) subdominant to the 3-body process.} For sufficiently small momentum transfers, we can describe the scattering process using chiral perturbation theory. In this approximation, the pions would scatter incoherently off the nucleons contained within the target's nuclei. Typically, staying within the regime of validity of chiral perturbation theory limits the mass of the BSM particles to $\lesssim 1 \ {\rm GeV}$, though for MCPs in particular, the production rate remains significant up to about 5 GeV. A full treatment of this process would require matching onto the perturbative QCD regime, which is beyond the scope of this work but has significant synergy with the DUNE neutrino cross section program~\cite{Ruso:2022qes}.

The total cross section for the scattering process in the lab frame is 
\begin{equation}
    \sigma = \frac{1}{4 E_\pi M_n} \int d \Pi \left (Z \langle \vert M^{(p)} \vert^2 \rangle 
    + (A-Z) \langle \vert M^{(n)} \vert^2 \rangle \right ),
    \label{eq:BSMCrosssection}
\end{equation}
where $E_{\pi}$ is the energy of the incoming pion, $M_n$ is the nucleon mass, $d \Pi$ is the 3- or 4-body phase space measure, $Z = 6$ is the atomic number of the graphite target, $A = 12$ is the mass number, and $\langle \vert M^{(p,n)} \vert^2 \rangle$ are the spin-averaged matrix elements for $\pi^{\pm}$ scattering off protons or neutrons, respectively.\footnote{There is a subtlety in treating the target mass in the cross section calculation. Here, we have treated the pions as scattering off individual nucleons, but the full center-of-mass energy (and hence the kinematic threshold) of the scattering process involves the nucleus mass. To treat the target mass properly, we should apply a target mass correction~\cite{Ruiz:2023ozv}, but in this work we have chosen to take the conservative approach (with the lowest possible kinematic threshold) and use the nucleon mass only.} 
For both axion and MCP production, we perform a Monte Carlo integral of the cross section using \texttt{MadGraph5}~\cite{Alwall:2014hca} to generate events. To remain within the regime of validity of leading-order chiral perturbation theory, we implement momentum cutoffs,
\begin{align}
\label{eq:tn}
t_n & \equiv -(p'_n - p_n)^2 \leq (4 \pi f_{\pi})^2 \\
\label{eq:tpi}
t_\pi & \equiv -(p'_\pi - p_\pi)^2 \leq (4 \pi f_{\pi})^2
\end{align}
as phase space cuts, where $p_n \ (p_\pi)$ and $p'_n \ (p'_\pi)$ are the initial and final nucleon (pion) 4-vectors, respectively. Because MCP production proceeds through a virtual photon, low-mass MCPs will be dominantly produced in the region of small photon virtuality, where the soft singularity of QED makes the Monte Carlo sampling unreliable. Consequently, we will focus only on MCPs above about 400 MeV, and we will use the closely-related study of dark photon production in~\cite{Curtin:2023bcf} to argue that there are limited sensitivity gains for lighter MCPs.

The cross section depends on the mass $m_X$ of the BSM particle $X$, its couplings to the SM generically denoted by $\kappa$, and the energy of the incoming pion $E_\pi$. We bin the charged pion energies into $i = 1,...,N_{0}$ bins, denoted $E_{\pi,i}$, and denote the number of charged pions in each energy bin as $N_{\pi, i}$. We then generate the cross section, $\sigma(E_{\pi,i}, m, \kappa)$, in \texttt{MadGraph5} for each energy bin, mass, and coupling value we consider. With the discretized charged pion energy distribution and cross section, we can calculate the number of BSM particles that can be produced at DUNE. Conservatively, we will restrict production to take place within the first interaction length of the pion, which is $l_\pi = 53.30$ cm for a graphite target. Since charged pions typically travel several interaction lengths before decaying and/or capturing on nuclei, additional production is possible deeper in the target, and is likely significant for lower-mass BSM particles; we intend to return to this point in a future in-depth study. For a given mass and coupling, the number of BSM particles produced within one interaction length of the dump is 
\begin{equation}
    N_X(m_X,\kappa) = L \times \sum_{i=1}^{N_0} N_{\pi, i} \,  \sigma_{\pi}(E_{\pi,i}, m_X, \kappa).
    \label{eq:Nx}
\end{equation}
Here the luminosity is defined as 
\begin{equation}
    L = l_\pi n_T \text{POT},
\end{equation}
where $n_T \approx 1.0 \times 10^{23}/{\rm cm}^{-3}$ is the target number density.

To determine the flux of these particles at DUNE ND, we only consider the particles that are within the angular resolution of the detector of interest at the DUNE ND Hall. We estimate this by generating $N_{\rm MC} = 10000$ Monte Carlo events for each BSM process and keeping only events that enter the detector geometry. The number of accepted BSM particles is then 
\begin{equation}
    N_X(m,\kappa)_{\rm acc.} = \frac{L}{N_{\rm MC}}\sum_{\rm events \in geom.} \sum_{i=1}^{N_0} N_{\pi, i} \,  \sigma_{\pi}(E_{\pi,i}, m, \kappa).
    \label{eq:numberflux}
\end{equation}

\subsection{BSM Particle Detection}
\label{subsec:BSMDetection}

Lastly, we must determine if the BSM particle produces a signal event in the detectors. If the BSM particle does not need to decay in order to be detected, as is the case for MCPs, the detection efficiency is simply a function of the detector properties, and we will explore this scenario in Sec.~\ref{sec:millicharged} below.

For heavy axions, detection relies on observing the decay products. We will only consider the long-lived case, where the BSM particle decays within the detectors at the DUNE ND Hall. The decay region consists of the length of LArTPC and MPD, which is 10 m long. We assume that if scintillators to detect MCPs are installed, these will not greatly alter this geometry. To calculate the number of signal events, we must determine the probability that the particle decays within the DUNE ND decay region,
\begin{equation}
    P_{\rm decay} = e^{-L_{\rm det}/(\gamma c \tau)} \times (1 - e^{-L_{\rm dec}/(\gamma c \tau)}),
    \label{eq:probdecay}
\end{equation}
where $L_{\rm det}$ is the distance to the detectors, $L_{\rm dec}$ is the decay region of the detectors, $\gamma$ is the boost factor of the BSM particle, and $\tau$ is its lifetime. The number of signal events is then
\begin{equation}
    N_{S} = P_{\rm decay} \times \text{BR}(X \to Y) \times N_{X, \rm acc.},
    \label{eq:signalevents}
\end{equation}
where $N_{X, \rm acc.}$ is given in Eq.~(\ref{eq:numberflux}) and $\text{BR}(X \to Y)$ is the branching ratio of the BSM particle $X$ decaying to the SM particles $Y$ which can be detected.

\section{Millicharged Particles}
\label{sec:millicharged}

To make contact with previous studies of MCPs~\cite{Kelly:2018brz,Harnik:2019zee}, we consider a model where a fermionic MCP $\chi$ with mass $m_\chi$ has electric charge $Q = e \epsilon$, with $e = \sqrt{4 \pi \alpha}$ the QED gauge coupling and $\epsilon \ll 1$. The MCP therefore couples to the SM photon $A_\mu$ through the Lagrangian
\begin{equation}
\mathcal{L}_{\chi \bar{\chi}} \supset \bar{\chi} \left( i \gamma^\mu \partial_\mu - m_\chi \right) \chi + Q \bar{\chi} \gamma^\mu \chi A_\mu + \mathcal{L}_{\pi N} + \mathcal{L}_{\pi N A},
\end{equation}
where $\mathcal{L}_{\pi N}$ and $\mathcal{L}_{\pi N A}$ are pion-nucleon and pion-nucleon-photon Lagrangians described by chiral perturbation theory. Following Refs.~\cite{Scherer:2002tk,Shin:2022ulh,Ellis:1997kc,Bernard:1995gx,Bernard:1997tq,Curtin:2023bcf}, pion-nucleon interactions can be described by
\begin{equation}
\begin{aligned}
\mathcal{L}_{\pi N} &\supset \frac{g_A}{2f_\pi} \biggr ( \bar{p} \gamma^\mu \gamma^5 p \partial_\mu \pi^0 - \bar{n} \gamma^\mu \gamma^5 n \partial_\mu \pi^0 
\\ &+ \sqrt{2} \bar{p} \gamma^\mu \gamma^5 n \partial_\mu \pi^+ + \sqrt{2} \bar{n} \gamma^\mu \gamma^5 p \partial_\mu \pi^- \biggr )
\\ &+ \frac{i}{4 f_\pi^2} \biggr ( \bar{p} \gamma^\mu p (\pi^+ \partial_\mu \pi^- - \pi^- \partial_\mu \pi^+) 
\\ &+ \bar{n}\gamma^\mu n (\pi^- \partial_\mu \pi^+ - \pi^+ \partial_\mu \pi^-)
\\ &+ \sqrt{2}\bar{n}\gamma^\mu p (\pi^- \partial_\mu \pi^0 - \pi^0 \partial_\mu \pi^-)
\\ &+ \sqrt{2} \bar{p} \gamma^\mu n (\pi^0 \partial_\mu \pi^+ - \pi^+ \partial_\mu \pi^0) \biggr )
\end{aligned}
\end{equation}
where $p$ and $n$ are the proton and neutron spinors, $\pi^0$ and $\pi^\pm$ are the pion fields, $f_\pi = 93$ MeV is the pion decay constant, and $g_A = 1.27$ is the axial coupling determined from neutron beta decay \cite{Czarnecki:2018okw}. Similarly, the pion-nucleon-photon interactions are described by
\begin{equation}
\begin{aligned}
\mathcal{L}_{\pi N A} &\supset i e A_\mu \biggr ( \pi^- \partial^\mu \pi^+ - \pi^+ \partial^\mu \pi^- \biggr )  
\\ &+ e^2 A_\mu A^\mu \pi^+ \pi^- + e \epsilon A_\mu \bar{p} \gamma^\mu p
\\ &+ \frac{i e g_A}{\sqrt{2} f_\pi} A_\mu \biggr (\bar{n} \gamma^\mu \gamma^5 p \pi^- - \bar{p} \gamma^\mu \gamma^5 n \pi^+ \biggr )
\\ &+ \frac{e}{2 f_\pi^2} A_\mu \biggr (\bar{n} \gamma^\mu n \pi^+ \pi^- - \bar{p} \gamma^\mu p \pi^+ \pi^-
\\ &+ \frac{1}{\sqrt{2}} \bar{n} \gamma^\mu p \pi^0 \pi^- + \frac{1}{\sqrt{2}} \bar{p} \gamma^\mu n \pi^0 \pi^+ \biggr ).
\end{aligned}
\end{equation}

\begin{figure*}
    \centering
    \includegraphics[width=0.48\textwidth]{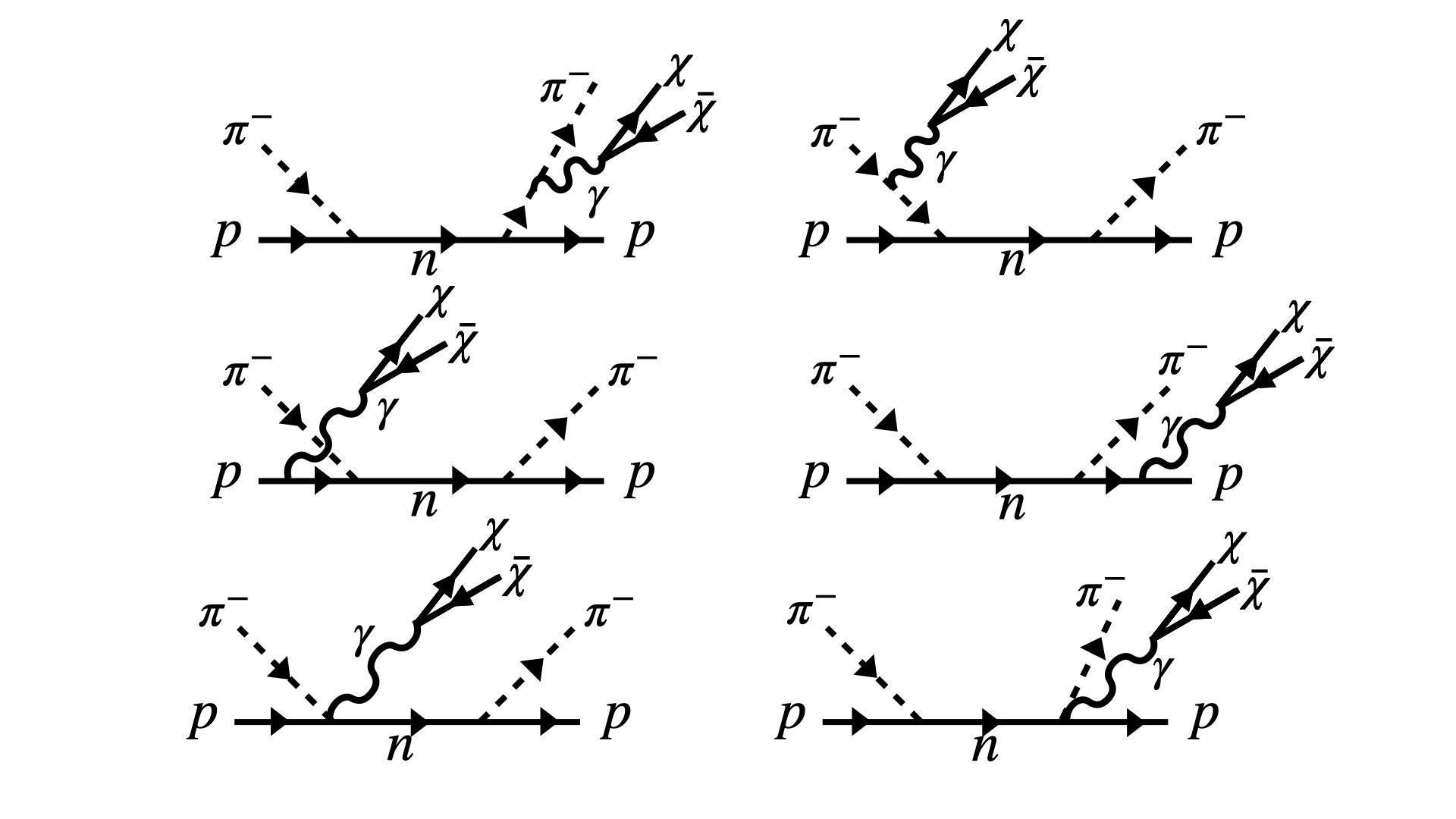}
    \includegraphics[width=0.48\textwidth]{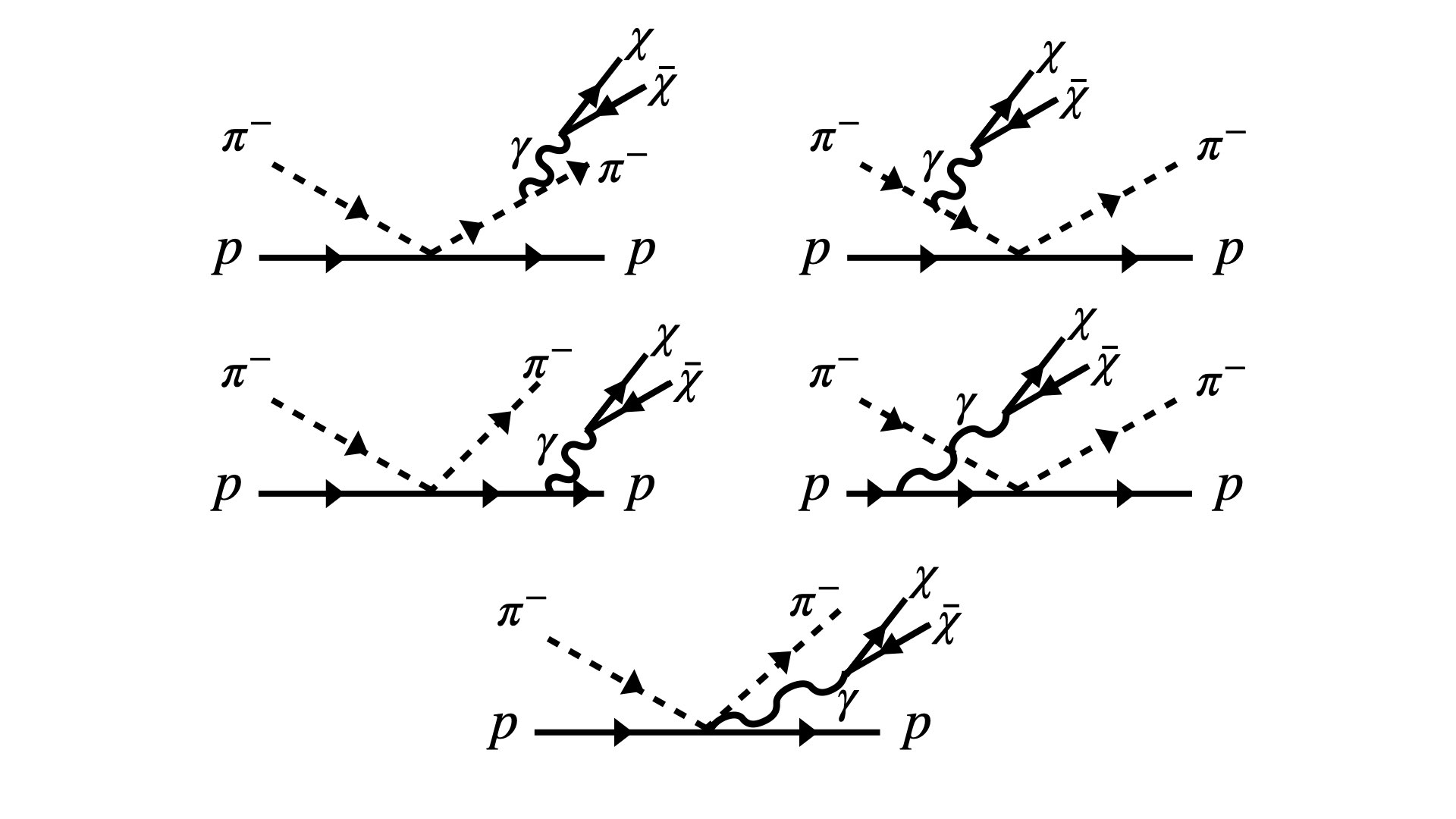}
    \caption{Leading-order Feynman diagrams for MCP production through $\pi^-$-proton scattering. A similar set of $t$-channel diagrams exists for $\pi^+$-proton scattering. Similarly, corresponding $s$- and $t$-channel diagrams also exist for $\pi^-$-neutron scattering.}
    \label{fig:feynman_diags1}
\end{figure*}

Our signal process is the scattering process $\pi^\pm N \to \pi^\pm N \chi \bar{\chi}$, where the $\chi \bar{\chi}$ pair is produced from a virtual photon radiated by the charged pion or the proton. 
The leading-order Feynman diagrams for MCP production from $\pi^- p$ scattering are shown in Fig.~\ref{fig:feynman_diags1}. Note that in order to preserve the Ward identity, we must sum over all diagrams and cannot claim that a subset are parametrically suppressed compared to the rest; the usual intuition that bremsstrahlung is suppressed for radiation by heavy particles does not hold in the non-renormalizable chiral Lagrangian~\cite{Curtin:2023bcf}. A similar set of diagrams related by crossing and isospin symmetry exist for $\pi^- n$, $\pi^+ p$, and $\pi^+ n$ scattering. To obtain the full cross section, we must sum incoherently over all four channels.

\subsection{MCP Flux}
\begin{figure}
    \centering
    \includegraphics[width=0.45\textwidth]{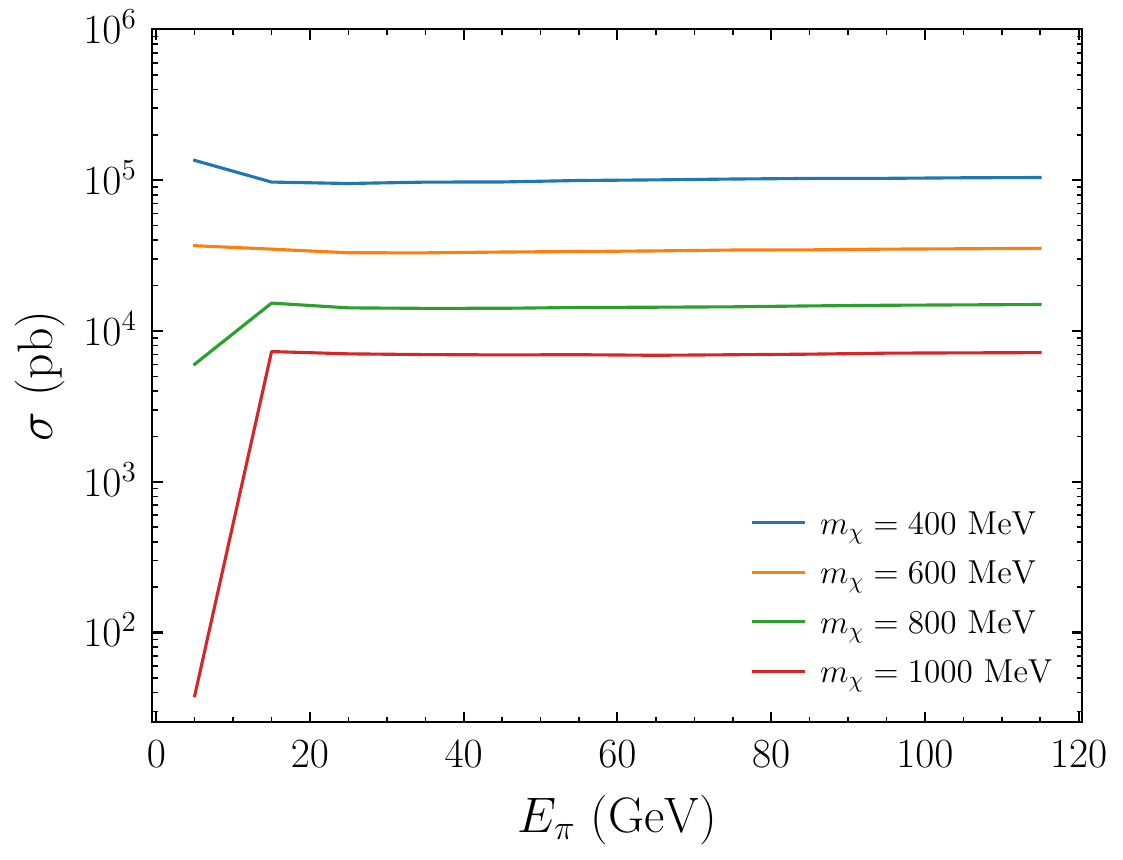}
    \caption{The total inclusive cross section for the scattering process $\pi N \to \pi N \chi \bar{\chi}$, summed over all charged pion and nucleon channels as in Eq.~(\ref{eq:BSMCrosssection}), as a function of incoming pion energy $E_{\pi}$ for various MCP masses $m_\chi$. The apparent increase in the cross section at low $E_\pi$ for $m_\chi = 400 \ {\rm MeV}$ is an artifact of the soft singularity of QED in \texttt{MadGraph}, which is why we restrict our analysis to $m_\chi > 400 \ {\rm MeV}$.}
    \label{fig:MCPCrossSec}
\end{figure}

 Given the large multiplicity of Feynman diagrams contributing to $\pi N \to \pi N \chi \bar{\chi}$ and the 4-body final state, we use \texttt{MadGraph} to calculate the cross section with momentum cutoffs given by Eqs.~(\ref{eq:tn})--(\ref{eq:tpi}) that ensures our cross sections are reliable within the regime of validity of chiral perturbation theory. As shown in Fig.~\ref{fig:MCPCrossSec}, the cross section is only weakly dependent on incident pion energy above the kinematic threshold for $\chi \bar{\chi}$ production.

\begin{figure}
    \includegraphics[width=0.48\textwidth]{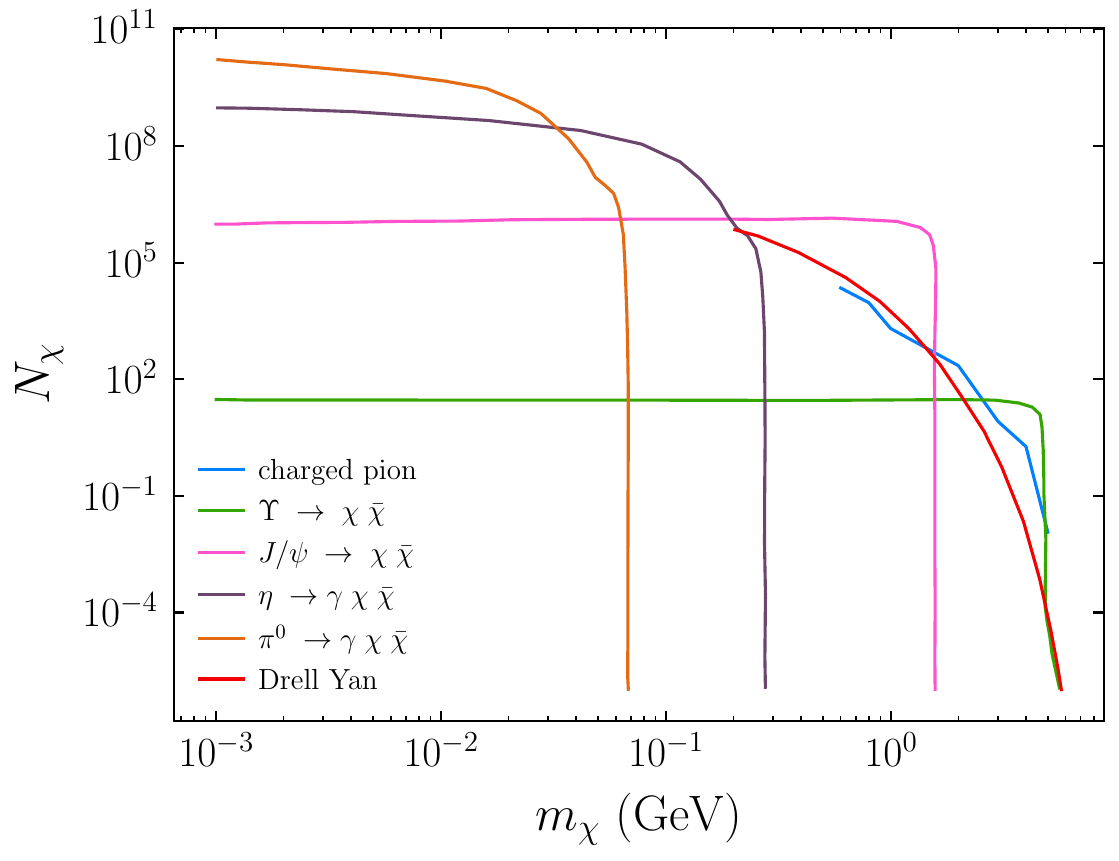}
    
    \caption{Number of MCP $N_\chi$ incident at the DUNE ND, for $10^{21}$ POT and $\epsilon = 10^{-3}$. Charged pion scattering exceeds Drell-Yan scattering for $m_\chi \gtrsim 1 \ {\rm GeV}$ up to highest masses ($\sim 5 \ {\rm GeV}$) allowed by the chiral cutoff, and is the dominant production mode between 2--3 GeV.}
    \label{fig:MCP_Flux}
\end{figure}

 The flux of MCP at the DUNE ND is given by 
 \begin{equation}
     N_{\chi} = L  N_{\pi^\pm} \left(\sigma_{\pi^+} + \sigma_{\pi^-}\right).
     \label{eq:Nchi}
 \end{equation}
To model how many MCPs can reach the detector, we implemented an angular cut on MCP trajectories as depicted in Fig.~\ref{fig:DUNE ND Complex}. For an expected detector area of $1 \ {\rm m}^2$ at a distance of 574 m from the target, the angular acceptance is  $\theta_{\rm acc.} = 8.71 \times 10^{-4} \; \text{rad}$. The expected flux of MCPs as a function of MCP mass is shown in Fig. \ref{fig:MCP_Flux}. For $m_\chi \gtrsim 1 \ {\rm GeV}$, we see that the MCP flux from charged pion production exceeds that of Drell-Yan production all the way up to the kinematic threshold of the 120 GeV beam. Interestingly, this is true even for the conservative assumption of the target mass equal to the bare nucleon mass, while for dark photon production in Ref.~\cite{Curtin:2023bcf}, the relative benefit of charged pion scattering is somewhat smaller. As alluded to earlier, \texttt{MadGraph} becomes unreliable for small MCP masses as the virtuality of the photon becomes small (this can be seen in the spurious increase of the cross section for $m_\chi = 400 \ {\rm MeV}$ as $E_\pi \to 0$ in Fig.~\ref{fig:MCPCrossSec}), so we do not show the flux for $m_\chi < 400 \ {\rm MeV}$. That said, based on the results of Ref.~\cite{Curtin:2023bcf} for similar kinematics, we strongly expect that the flux will be subdominant to meson decays in this mass range and will not yield any additional gains in sensitivity.

\subsection{MCP Detector Signatures and Backgrounds}
\label{sec:MCP_Signature}

The proposed MCP detector design is based on that of MilliQan \cite{Haas:2014dda, Ball:2016zrp}. The signature for detecting a MCP is three soft ionizations through the scintillator arrays that are collinear with the beam line and target, detectable through the collection of photoelectrons by PMTs. Following Ref.~\cite{Kelly:2018brz}, we also require that all three soft ionizations occur within a small 15 nanosecond window to further reduce background.

There are two types of background that would look like additional soft ionizations throughout the scintillators: detector-related background and beam-related background. To make a direct comparison with Ref.~\cite{Kelly:2018brz}, we make identical assumptions about the backgrounds, yielding an estimate of about 600 events per year; we refer the reader to Ref.~\cite{Kelly:2018brz} for more details about the background processes.

\subsection{MCP Sensitivity at Dune ND}
In the FerMINI design \cite{Kelly:2018brz,ParticleDataGroup:2016lqr}, a minimum-ionizing particle with charge $|Q| = 1$ would produce approximately $10^6$ photoelectrons (PE) in a 1 m plastic scintillator with a density of $\rho_{\rm scint.} = 1$ g/cm$^3$. Thus MCPs with charge $Q = e \epsilon$ would produce PEs that scale as $\epsilon^2$. We can then quantify the probability of a MCP producing at least one PE in all three scintillator stacks as 
\begin{equation}
P(\epsilon) = \left(1-e^{-\overline{N}_{ \rm PE}(\epsilon)}\right)^3,
\label{eq:prob}
\end{equation}
where $\overline{N}_{\rm PE}$ denotes the average number of photoelectrons produced from soft ionization. We use the approximation of
$\overline{N}_{\rm PE}(\epsilon) = \left(\frac{\epsilon}{\zeta} \right)^2$
where $\zeta = 2 \times 10^{-3}$. This value of $\zeta$ corresponds to a Saint-Gobain BC-408 plastic scintillator \cite{Kelly:2018brz, Ball:2016zrp}. Given these choices, such an experiment could be sensitive to $\epsilon$ of order $10^{-4}$.

\begin{figure}[t]
    \centering
    \includegraphics[width=0.5\textwidth]{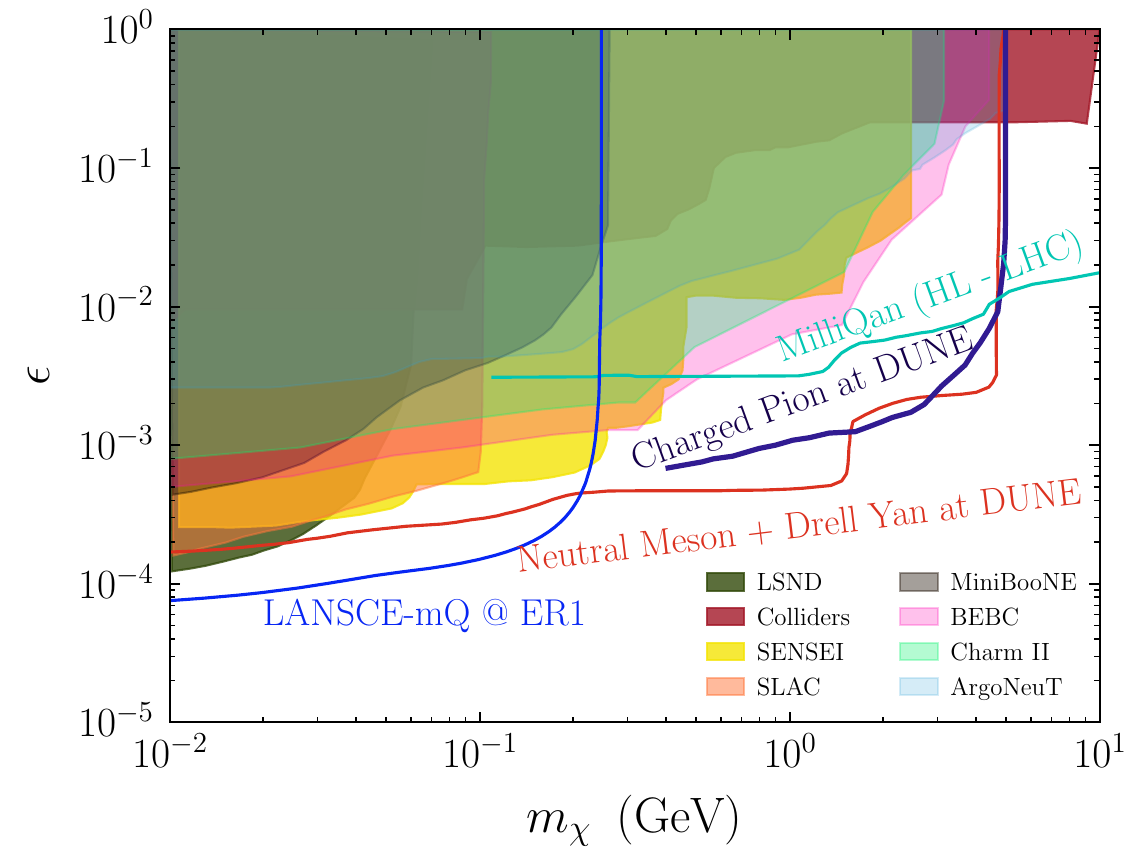}
    \caption{95\% CL sensitivity to MCP detection at DUNE ND with $10^{21}$ POT (about one year of beam time). The charged pion production channel is shown in purple, and the red-orange curve labeled ``Neutral Meson + Drell Yan'' is our reproduction of the projected sensitivity from Ref.~\cite{Kelly:2018brz}. The milliQan HL-LHC projected sensitivity~\cite{Haas:2014dda,Ball:2016zrp} is plotted in teal, and the LANSCE-mQ at ER1 \cite{Tsai:2024wdh} projected sensitivity, using $5.9 \times 10^{22}$ POT (about 3 years of beam time), is plotted in blue.  Shaded regions are existing limits from SLAC (\cite{Prinz:2001qz}, peach), SENSEI (\cite{SENSEI:2019ibb}, yellow), BEBC (\cite{Marocco:2020dqu}, light pink), Charm II (\cite{CHARM-II:1989nic}, light green), LSND (\cite{Magill:2018tbb}, olive green), MiniBooNE (\cite{Magill:2018tbb}, gray), ArgoNeuT (\cite{ArgoNeuT:2019ckq}, light blue), and colliders (\cite{Davidson:2000hf}, burgundy).}
    \label{fig:chargedPionReach}
\end{figure}

With the probability of detecting a PE in the scintillators, we can now calculate the number of MCP signal events using Eqs.~(\ref{eq:Nchi}) and~(\ref{eq:prob}),
\begin{equation}
N_S = P(\epsilon) \, N_\chi (m_\chi, \epsilon).
\end{equation}
Assuming that the scintillators can detect values as low as $\epsilon = 10^{-4}$, we compute the projected 95\% CL limit, i.e. $N_S/\sqrt{N_B} = 2$ where $N_B$ is the number of background events, for the DUNE ND setup to MCP from charged pion scattering. The results are shown in Fig.~\ref{fig:chargedPionReach} along with constraints and projections from other beam-dump, collider, and direct detection experiments. We also show the projected limit from neutral meson decay and Drell-Yan scattering at DUNE, following Fig.~1 of Ref.~\cite{Kelly:2018brz} for the total number of MCPs that reach the target, and then applying Eq.~(\ref{eq:prob}) in order to model the scintillator response. We find that we reproduce the sensitivity in Ref.~\cite{Kelly:2018brz} to within about a factor of 2, with any differences likely due to slightly different treatments of the angular acceptance~\cite{KellyPrivate}, which lets us directly assess the increase in sensitivity from charged pion scattering. Overall, the charged pion production channel (which is irreducible in this model) can offer additional sensitivity to detecting MCPs for masses $1.5 \ {\rm GeV} \lesssim m_\chi \lesssim 3 \ {\rm GeV}$.

\section{Heavy Axion-like Particles}
\label{sec:axions}

The DUNE ND is an excellent experiment to look for long-lived exotic particles at MeV-GeV mass scales. One such particle that we can search for at DUNE ND is a heavy axion coupled to gluons at low energies. This particle may be a generic ALP, but in some models with additional gauge groups, such a heavy axion may in fact be the QCD axion which solves the strong CP problem~\cite{Hook:2019qoh}. In this section, we will be agnostic as to the UV interpretation of such a particle, and simply use this model as a benchmark to make a direct comparison with the model of Ref.~\cite{Kelly:2020dda}. As we have done with MCPs in Sec.~\ref{sec:millicharged}, we will demonstrate the additional sensitivity gained from charged pion production, which is an irreducible production channel in this model.

As in Ref.~\cite{Kelly:2020dda}, we consider a low-energy ALP model where at scales $\mu=\mathcal{O}(1)$ GeV the ALP is only coupled to gluons,
\begin{equation}
\begin{split}
    \mathcal{L}_a =& \, \frac{\alpha_s}{8\pi f_a} a \, G_{\mu \nu}^a \tilde{G}^{a \mu \nu}.
\end{split}
\end{equation}
Here, $f_a$ is the axion decay constant, $\alpha_s$ is the strong fine structure constant, and $G^a_{\mu \nu}$ is the gluon field strength tensor. Following~\cite{DiLuzio:2020wdo,Chang:1993gm,Choi:2021ign}, the axion couplings to nucleons and pions in the chiral Lagrangian can be derived by performing a chiral rotation to shift the axion field into the quark mass matrix. This rotation is arbitrary, but in the conventional parameterization which eliminates mass mixing between the axion and the $\pi^0$, we obtain
\begin{equation}
    \begin{split}
    \mathcal{L}_{a\pi N} =& \frac{g_A}{2 f_\pi} \bigg( \partial_\mu \pi^0 (\bar{p} \gamma^\mu \gamma_5 p - \bar{n} \gamma^\mu \gamma_5 n) 
    \\&+ \sqrt{2}\partial_\mu \pi^+ \bar{p} \gamma^\mu \gamma_5 n + \sqrt{2}\partial_\mu \pi^- \bar{n} \gamma^\mu \gamma_5 p \bigg)
    \\&+ \frac{\partial_\mu a}{2 f_a} \bigg( C_{ap} \bar{p} \gamma^\mu \gamma_5 p + C_{an} \bar{n} \gamma^\mu \gamma_5 n 
    \\&+
    \frac{C_{a \pi N}}{f_{\pi}}(i \pi^+ \bar{p} \gamma^\mu n - i \bar{n} \gamma^\mu p) \bigg)
    \\+ \frac{\partial_\mu a}{2 f_a} \frac{C_{a \pi}}{f_\pi} ( &\pi^0 \pi^+ \partial^\mu \pi^- + \pi^0 \pi^- \partial^\mu \pi^+ - 2 \pi^+ \pi^- \partial^\mu \pi^0 ), 
    \end{split}
    \label{eq:axionLagr}
\end{equation}
where the various axion couplings are given as
\begin{equation}
    \begin{split}
        C_{ap} - C_{an} &= g_A \left( \frac{m_u - m_d}{m_u + m_d}\right), \\
        C_{ap} + C_{an} &= -g_0, \\
        C_{a \pi N} &= \frac{C_{ap} - C_{an}}{\sqrt{2} g_A}, \\
        C_{a \pi} &= \frac{2 (C_{ap} - C_{an})}{3 g_A}.
    \end{split}
\end{equation}
As in Sec.~\ref{sec:millicharged}, $f_\pi = 93$ MeV is the pion decay constant, $g_A = 1.27$ is the nucleon axial coupling, and $g_0 = 0.52$ is another axial coupling extracted from lattice QCD simulations and evaluated at an $\overline{\rm MS}$ scale $\mu = 2$ GeV \cite{GrillidiCortona:2015jxo}. Note that the chiral rotation which generates these couplings will also generate an axion-photon coupling; in the spirit of this ``gluon dominance'' model, we will ignore production from this operator and only use it to permit decays $a \to \gamma \gamma$ for $m_a < 3m_\pi$ (see Sec.~\ref{sec:ALPDetector} below). Furthermore, we note that axion-neutral meson kinetic mixing will also be present in the chiral Lagrangian. In order to compare the parametric dependence of scattering processes to previously-considered mixing processes, we are neglecting diagrams which involve kinetic mixing insertions, and we leave to future work a full analysis which simultaneously treats mixing and scattering in order to capture any potential interference effects.

\subsection{ALP Flux}
\label{sec:axion flux}
\begin{figure*}
    \centering
    \includegraphics[width=0.6\textwidth]{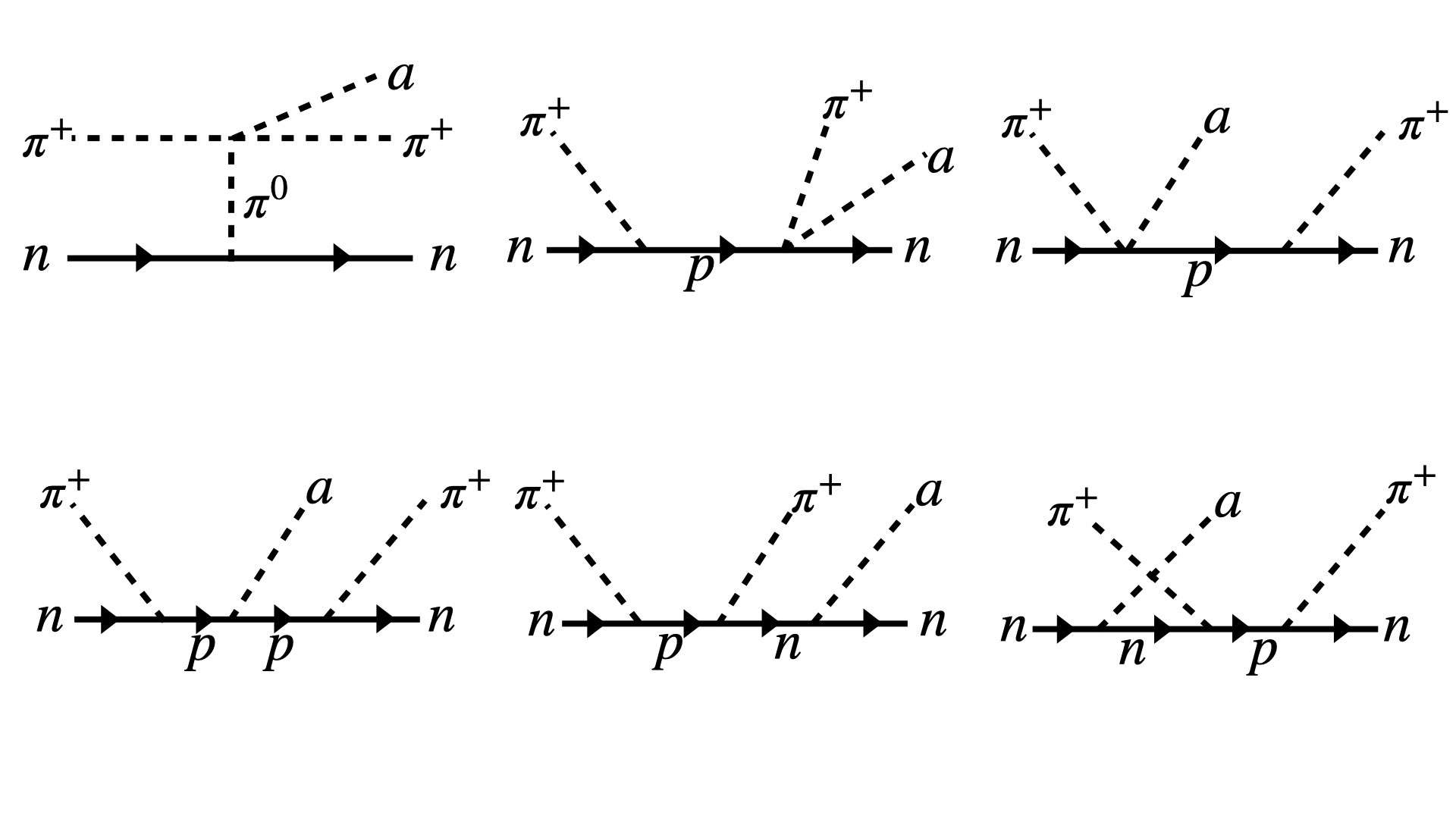}
    \caption{Leading-order diagrams for axion production from $\pi^+$-neutron scattering. A similar set of $t$-channel diagrams exists for $\pi^+$-proton scattering, and corresponding $s$- and $t$-channel diagrams also exist for $\pi^-$ scattering.}
    \label{fig:axionFeynmanDiagrams}
\end{figure*}

\begin{figure}
    \centering
    \includegraphics[width=0.45\textwidth]{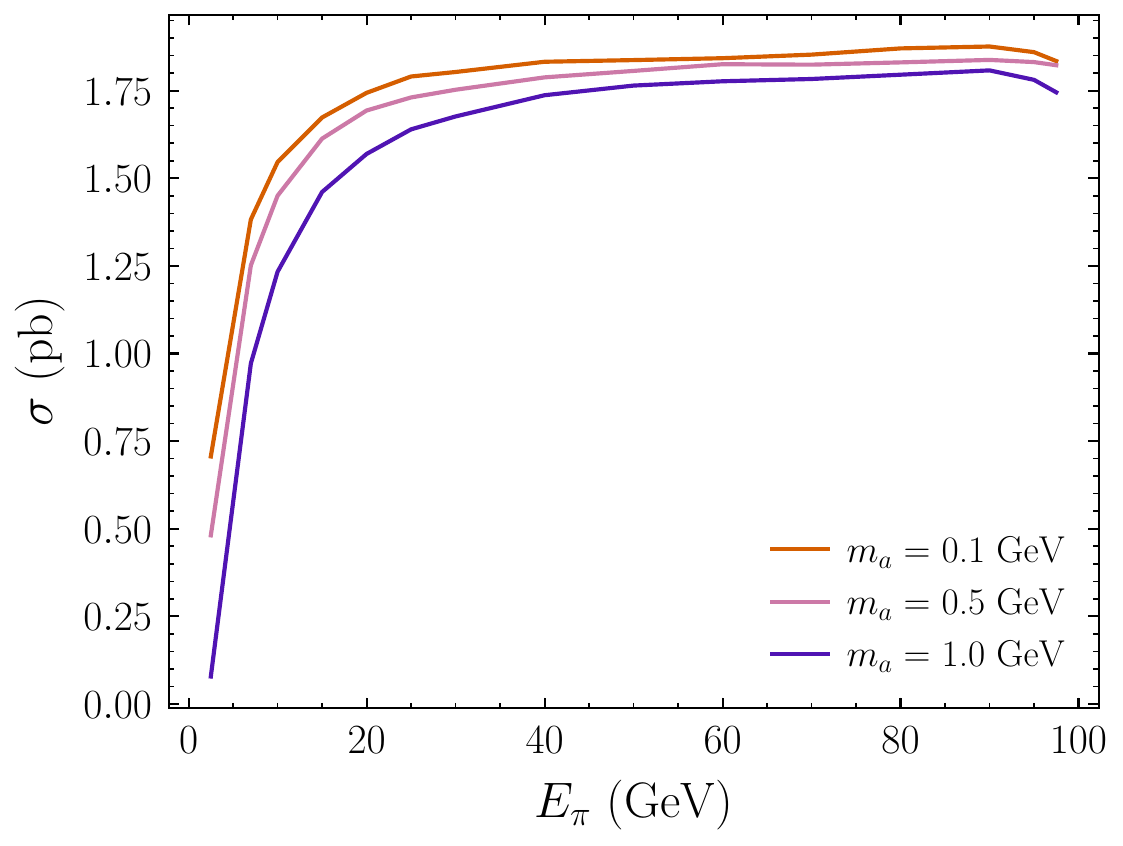}
    \caption{The cross section for the scattering process $\pi N \to \pi N a$ as a function of incoming pion energy $E_{\pi}$ for range of axion masses and $f_a = 10^4$ GeV. The cross section scales with increasing pion energy due to the derivative couplings in the chiral Lagrangian. The cross section also has a weak dependence on $m_a$, which is only significant at low pion energies.}
    \label{fig:axionCrossSec}
\end{figure}

Axions may be emitted from pion-nucleus scattering through $\pi^{\pm} N \to \pi^{\pm} N' a$. Below squared 4-momentum transfers of $(4 \pi f_{\pi})^2$, we can describe the scattering using chiral perturbation theory, using the Lagrangian in (\ref{eq:axionLagr}). The pions would incoherently scatter off the nucleons contained within the target nuclei. The leading-order amplitudes Feynman diagrams for $\pi^+ n$ scattering are shown in Fig. \ref{fig:axionFeynmanDiagrams}, with similar diagrams for the other charge and isospin channels. We implement this model in \texttt{Madgraph} according to the methods in Sec ~\ref{subsec:chargedpionproduction}, and show the cross section as a function of incident pion energy in Fig.~\ref{fig:axionCrossSec}. The derivative couplings in Eq.~(\ref{eq:axionLagr}) yield a cross section growing faster with $E_\pi$ than the MCP cross section studied in Sec.~\ref{sec:millicharged}. In addition, we note that our cross section has very weak $m_a$ dependence until the kinematic threshold imposed by the chiral momentum cutoff.

Since the axion has the same quantum numbers as the $\pi^0$, axions may also be produced through conversion by replacing the $\pi^0$ in any isospin-exchange process, e.g.\  $\pi^+ n \to a p$. However, it turns out that the production cross section for $2 \to 3$ scattering is an order of magnitude larger than $2 \to 2$ scattering over all masses and couplings we consider. This is perhaps surprising given that one would naively expect the $2 \to 3$ process to be phase-space suppressed. However, charge conservation only allows a $\pi^+$ to scatter off a neutron and the $\pi^-$ to scatter off a proton in $2\to 2$ scattering. In contrast, in the $2 \to 3$ process, a $\pi^+$ could scatter off both protons and neutrons, taking advantage of the full incoherent nuclear scattering. In addition, the $2 \to 3$ process contains six Feynman diagrams compared to three diagrams for $2 \to 2$, and the combinatoric addition of these diagrams seems to overcome the phase space suppression from an additional final-state particles. Furthermore, while the axion could also be produced from scattering initiated by neutral pions as well as charged pions, the neutral pion cross section is suppressed due to the particular values of the coupling constants $C_{ap}, C_{an}, \text{ and } C_{a \pi N}$ in the SM.\footnote{Due to the arbitrariness of the chiral rotation which generates the axion-pion couplings, and given that our choice eliminates axion-pion mass mixing, it is possible that the relative size of the $\pi^0$ cross section is basis-dependent, but the physical inclusive cross section is basis-independent. We thank Joshua Berger for discussions on this point.} We explore this somewhat counterintuitive feature of the axion cross section in Appendix \ref{app:axion_xsec}.

We can now calculate the flux of axions at the DUNE ND, using
\begin{equation}
    \Phi_a = \frac{L}{A} \times (N_{\pi^+} \sigma_{\pi^+} + N_{\pi^-} \sigma_{\pi^-}),
    \label{eq:Na}
\end{equation}
where $A$ is the cross-sectional area of the detector. As with MCPs, we will just compute the number of axions produced within the first pion interaction length in the target, and as such, we will not consider the effect of subsequent target layers on the pion energy distribution.

\begin{figure}
    \centering
    \includegraphics[width=0.48\textwidth]{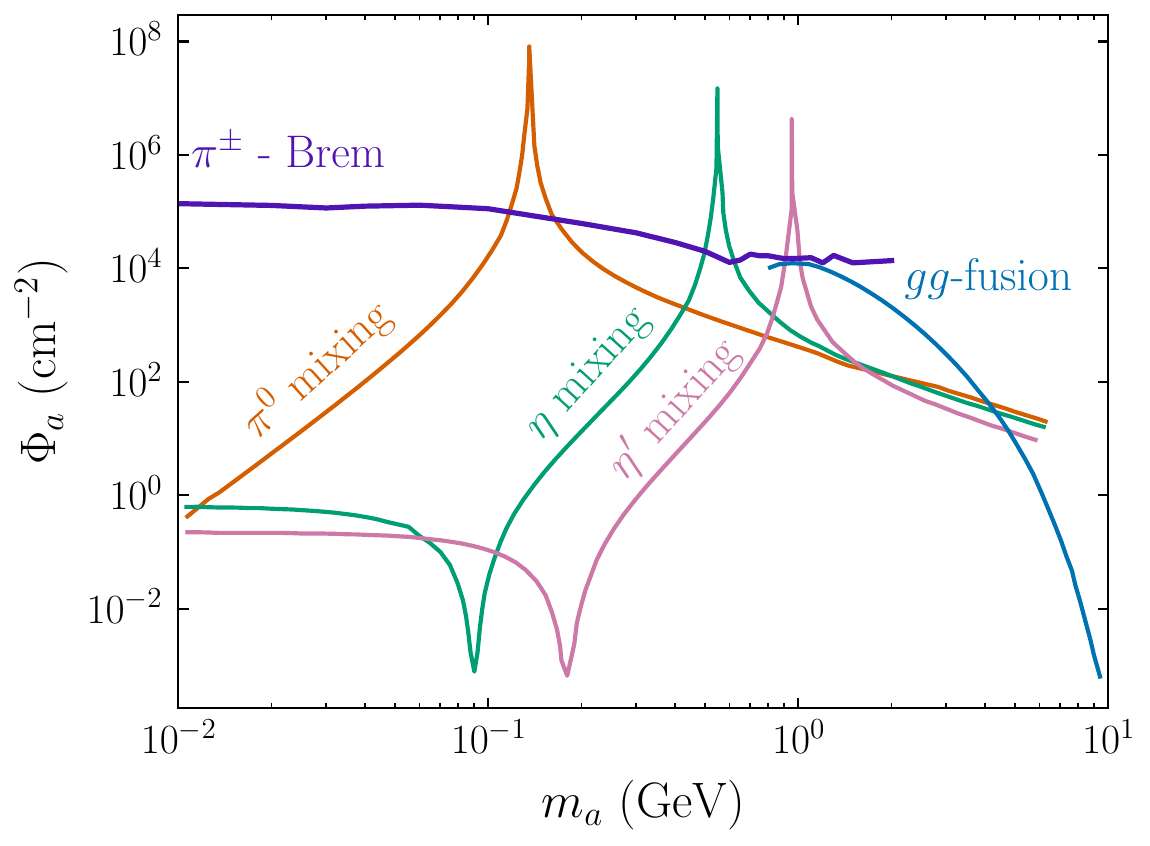}
    \caption{The expected flux of axions at ArgonCube and MPD, located at the DUNE ND Hall. We compare the axion production from charged pion bremsstrahlung (blue) to the production from $\pi^0$ decay (orange), $\eta$ decay (green), $\eta'$ decay (pink), and gluon-gluon fusion (purple), all assuming $f_a = 10^4$ GeV. For $m_a \lesssim 4\pi f_\pi$,  axion production from charged pion scattering dominates the flux, with the exception of where $m_a$ resonates with the neutral meson masses.}
    \label{fig:axionFlux}
\end{figure}

Fig.~\ref{fig:axionFlux} shows the flux of axions produced that reach ArgonCube and MPD at the DUNE ND Hall. We compare the flux from charged pion bremsstrahlung to the flux from neutral meson decay and gluon-gluon fusion. With the exception of the resonance peaks, the flux from charged pion bremsstrahlung is approximately $\mathcal{O}(10 - 10^6)$ larger than the flux from neutral meson decay and gluon-gluon fusion depending on the axion mass. This increase in flux will increase the sensitivity of the DUNE ND to axion detection for $m_a$ away from the neutral meson masses.

We can understand this enhancement through a dimensional analysis estimate. The number of axions produced through neutral pion mixing is~\cite{Kelly:2020dda}
\begin{equation}
    N_{\pi^0 \to a} \sim N_{\rm POT} N_{\pi^0} \left( \frac{1}{6} \frac{f_\pi}{f_a} \frac{m_a^2}{m_a^2-m_\pi^2} \right)^2.
\end{equation}
For the production of axions through charged pion scattering, each diagram in Fig.~\ref{fig:axionFeynmanDiagrams} has two pion lines contributing a factor of $1/f_\pi$ each, an axion contributing $1/f_a$, and derivative couplings which contribute factors of the overall energy scale $E_\pi$. The $2 \to 3$ cross section therefore scales as 
\begin{equation}
    \sigma(\pi N \to \pi N a) \sim \frac{1}{16\pi} \frac{E_\pi^4}{f_a^2 f_\pi^4},
\end{equation}
where we have included a 3-body phase space factor $\frac{1}{16\pi}$ and neglected the order-1 factors $g_0$ and $g_A$ as well as any combinatorial factors from the multiplicity of Feynman diagrams. The total number of axions will scale as $\sigma \times l_\pi n_T N_{\pi^\pm}\, {\rm POT}$, so the ratio of production channels assuming $m_a \ll m_\pi$ and $N_{\pi^\pm} \approx 2 N_{\pi^0}$ is
\begin{align}
\frac{N_{\pi N \to \pi N a}}{N_{\pi^0 \to a}} & \sim \frac{9}{2\pi}\frac{E_\pi^4}{m_a^4} \frac{l_\pi n_T m_\pi^4}{f_\pi^6} \nonumber \\
& \sim 1.5 \times 10^8 \left(\frac{E_\pi}{1 \ {\rm GeV}}\right)^4 \left(\frac{10 \ {\rm MeV}}{m_a}\right)^4.
\end{align}
This is reasonably close to the $\sim 10^6$ enhancement seen in Fig.~\ref{fig:axionFlux}, which accounts for all of the order-1 coupling factors, the true pion energy distribution, and the detector's geometric acceptance.

\subsection{ALP Detector Signatures and Backgrounds}
\label{sec:ALPDetector}

The axions that are created within the angular acceptance of ArgonCube and MPD at DUNE ND can be detected by their decay into SM particles. For the axion masses we are interested in, the axion can decay into photons or hadrons. Even in the gluon-dominated model, axions are able to decay to photons through mixing with neutral mesons (indeed, this was the primary production mechanism of Ref.~\cite{Kelly:2020dda}). For $m_a < 3 m_\pi$, the axion will primarily decay into two photons, $a \to \gamma \gamma$. The decay width is given by~\cite{Aloni:2018vki}
\begin{equation}
    \Gamma_{a \to \gamma \gamma} = \frac{\alpha_{\rm EM }}{256 \pi^3} \frac{m_a^3}{f_a^2}c^2_\gamma,
\end{equation}
where
\begin{equation}
    \begin{split}
        c_\gamma =& -1.92 + \frac{1}{3} \frac{m_a^2}{m_a^2 - m_\pi^2} + \frac{8}{9} \frac{m_a^2 - \frac{4}{9} m_\pi^2}{m_a^2 - m_\eta^2} \\
        &+ \frac{7}{9} \frac{m_a^2 - \frac{16}{9} m_\pi^2}{m_a^2 - m_\eta'^2}.
    \end{split}
\end{equation}
For $3 m_\pi < m_a \lesssim 1$ GeV, the axion decays to primarily to hadrons through $a \to 3 \pi$ and $a \to \pi \pi \gamma$ \cite{Aloni:2018vki, Kelly:2020dda}. 

We consider the case where the axion decays within the liquid and gaseous argon detectors at DUNE ND, between 574 and 584 m from the target, with an angular acceptance of $\sim$ 3 mrad (see Fig.~\ref{fig:DUNE ND Complex}). Approximately 1\% of the total number of axions produced are detected due to the small angular acceptance. These axions must also be highly boosted in order to travel 574 m and be within the angular resolution of the detectors. As a result, the decay products are also highly boosted compared to background events \cite{Kelly:2020dda}. Much of the background in both detectors comes from neutrino-related scattering processes in the argon tanks \cite{Berryman:2019dme, Kelly:2020dda}. Both the diphoton and hadronic backgrounds would have low energies and would result in a roughly isotropic signature. Since our axion signal is highly boosted, we can easily distinguish signal from background events~\cite{Kelly:2020dda}.

\subsection{ALP Sensitivity at DUNE ND}

Combining the expected signal from $a \to \gamma \gamma$ and $a \to$ hadrons, we can determine the axion parameter space to which DUNE ND is sensitive. We calculate the number of signal events according to the methods described in Sec.~\ref{subsec:BSMDetection}, using the axion decay length $c \tau_a$ from Refs.~\cite{Kelly:2020dda,Blinov:2021say}. The number of axion signal events is then
\begin{equation}
    N_{S} = P_{\rm decay} \times \text{BR}(a \to X) \times N_a,
\end{equation}
where $N_a = A \Phi_a$ with $\Phi_a$ in (\ref{eq:Na}), and $\text{BR}(a \to X)$ is the branching ratio of the axion decaying to either photons or hadrons~\cite{Aloni:2018vki, Blinov:2021say}. 

\begin{figure}
    \centering
    \includegraphics[width=0.48\textwidth]{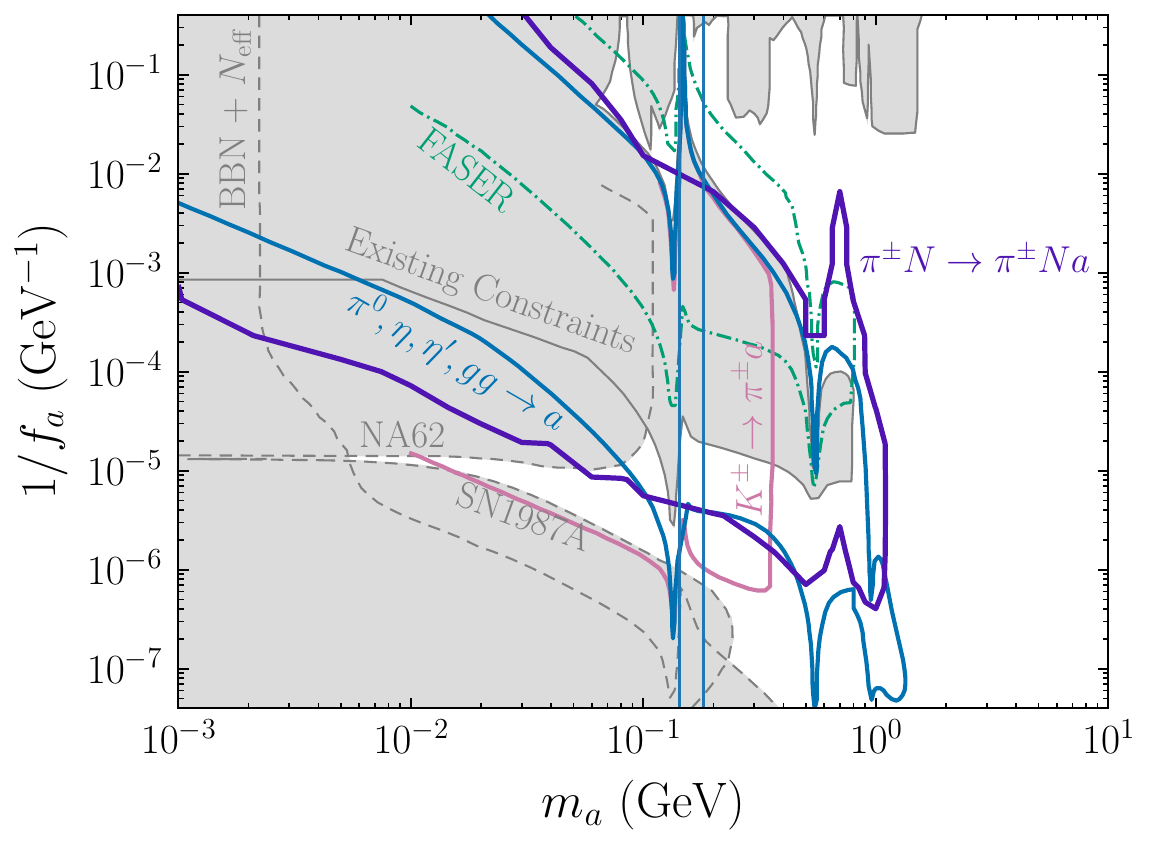}
    \caption{The sensitivity of DUNE ND to 3 or more heavy ALP signal events over 10 years of data collection from charged pion scattering (purple) along with other projections and constraints. We show projections for DUNE ND (solid) along with FASER \cite{FASER:2018eoc} (green dot-dashed). The projection for an axion signal from neutral meson and gluon-gluon fusion production are shown in blue \cite{Kelly:2020dda}. The pink curve, adapted from Ref.~\cite{Berger:2024xqk} and rescaled to the same event rate as the other curves, shows diphoton signal events from $K^\pm \to \pi^\pm a$. In shaded gray we show constraints from NA62 \cite{NA62:2021zjw} for invisible kaon decays, SN1987A \cite{Ertas:2020xcc, Chang:2018rso}, and cosmology \cite{Depta:2020wmr}. In the shaded region labeled ``Existing Constraints" we include bounds from invisible kaon decays from E787 and E949 \cite{Ertas:2020xcc}, electron beam dump \cite{Dobrich:2015jyk, Dolan:2017osp, NA64:2020qwq}, CHARM \cite{CHARM:1985anb}, visible kaon decays \cite{Gori:2020xvq}, $B$ meson decays \cite{Aloni:2018vki}, and LHC di-jet searches \cite{Mariotti:2017vtv}. }
    \label{fig:axionSensitivity}
\end{figure}

In Figure \ref{fig:axionSensitivity}, we show the sensitivity of DUNE ND to 3 or more signal events in 10 years of data collection. The increased production rate from charged pion scattering away from meson resonances improves the projected sensitivity of DUNE in both the long-lifetime (small $1/f_a$) and short-lifetime (large $1/f_a$) regimes. Because we only calculated $N_a$ up to the chiral cutoff $t_n, t_\pi < (4 \pi f_\pi)^2$, we do not find additional sensitivity beyond the largest $m_a$ considered in Ref.~\cite{Kelly:2020dda}. For axions below the kaon mass, the sensitivity from charged pion scattering alone is within an order-1 factor of axions produced from kaon decay, recently argued in Ref.~\cite{Berger:2024xqk} to be the dominant production channel in this mass range. Including charged pion scattering beyond the first interaction length of the dump could yield additional sensitivity, possibly making these two production modes comparable. 

\section{Conclusion}
\label{sec:conclusions}

In this work, we explored the sensitivity of DUNE ND to heavy axions and millicharged particles produced through charged pion scattering. We found that the expected flux of both ALPs and MCP produced from charged pion scattering was greater than previously-considered production modes meson for some masses in the MeV--GeV range. With this increased flux, we were able to claim a conservative increase in sensitivity of DUNE ND to both ALPs and MCPs. The increase in sensitivity to MCP was modest, but combined with the additional production mechanisms can modestly increase the sensitivity at masses above 2 GeV. The increase in sensitivity to ALPs is similar, and combined with the strong projected constraints from kaon decays in Ref.~\cite{Berger:2024xqk}, can likely fully probe remaining parameter space for $m_a \lesssim 250 \ {\rm MeV}$ between the recent NA62 limits and the SN1987A constraint.

For both sensitivity calculations, we only considered the BSM particle production from a scattering event within the first pion interaction length. However, the energy of the pion is not completely depleted after this event but is only reduced by a factor of $1/e$ on average.  The sensitivity to ALPs and MCPs could very likely be improved by incorporating multiple pion scatterings through the target. In addition, there is a substantial flux of pions that survive the $\sim 200$ m decay region after being focused toward the DUNE detectors, which could rescatter in the rock between the target and ND complex. Axions produced from these charged pion scattering events would only need to travel half the distance to the DUNE ND, thereby increasing the flux of short-lifetime ALPS at large $1/f_a$ which would not have otherwise survived to the detectors. We leave this extension to future work. 

\begin{figure}[t!]
    \centering
    \includegraphics[width=0.46\textwidth]{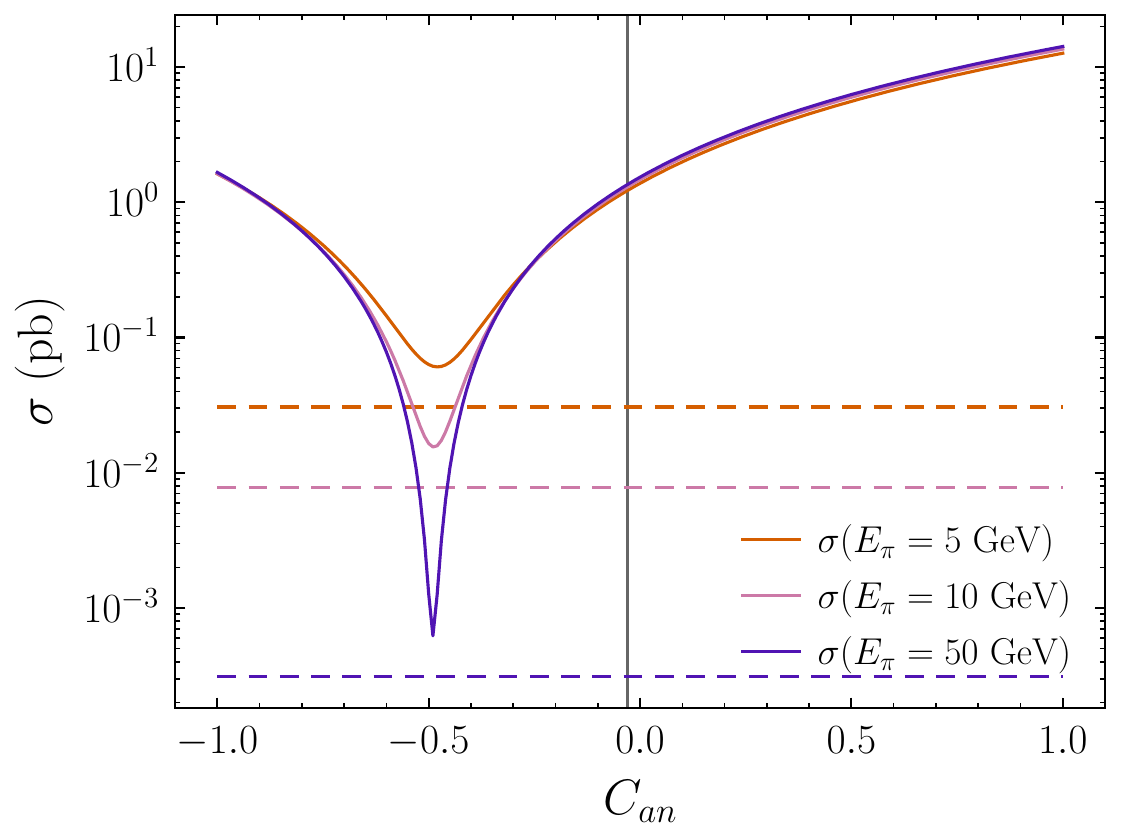}
    \caption{The cross sections $\sigma(\pi^- p \to a n)$ (solid) and $\sigma(\pi^0 p \to a p)$ (dashed) as a function of $C_{an}$ values. We used the parameters $m_a = 0.5$ GeV, $f_a = 10^4$ PeV, and $C_{ap} = -0.49$. The vertical gray line indicates our model value $C_{an} = -0.03$. We have plotted the cross section for three incoming pion energies $E_\pi = 5$ GeV (orange), $E_\pi = 10$ GeV (pink), and $E_\pi = 50$ GeV (purple). We see that $\sigma(\pi^- p \to a n)$ approaches $2 \times \sigma(\pi^0 p \to a p)$ as $C_{an}$ approaches $C_{ap}$. For all other values of $C_{an}$, $\sigma(\pi^- p \to a n)$ exceeds $\sigma(\pi^0 p \to a p)$ by at least a factor of 2.}
    \label{fig:varyCan_varyEpi}
\end{figure}

Finally, it is reassuring that our chiral perturbation theory calculation closely matches the parton-level Drell-Yan or gluon fusion calculations for BSM particle masses around the chiral cutoff, but there are likely additional gains in sensitivity to both MCPs and axions from a complete treatment of the high-momentum-transfer regime which interpolates between the chiral Lagrangian and perturbative QCD. Such a treatment would involve, for example, quasi-elastic scattering processes and target mass corrections, and for ALPs would consider the scattering and mixing processes in the same theoretical framework. This study would be highly synergistic with the DUNE neutrino physics program where the intermediate-energy regime is crucial for understanding neutrino-nucleus cross sections~\cite{Ruso:2022qes}. Further improvements in the theoretical framework for GeV-scale neutrino cross sections at DUNE would therefore translate directly into improved new physics sensitivity at DUNE, and we look forward to such collaborations involving lattice QCD, BSM phenomenology, and neutrino physics to maximize the science potential of DUNE.

\section*{Acknowledgments}
We thank Joshua Berger, Roni Harnik, Kevin Kelly, Zhen Liu, and Yu-Dai Tsai for helpful discussions. We also thank Insung Hwang for their work collecting constraints for MCP detection. We thank the CETUP* 2024 conference for facilitating discussions related to this work. This work is supported in part by DOE grant DE-SC0015655.

\appendix
\section{Axion cross section peculiarities}
\label{app:axion_xsec}

In this Appendix, we investigate some unusual features of the pion-nucleon-axion cross section. First, we consider the relative sizes of the charged and neutral pion cross sections. In our calculation of $\sigma(\pi^0 N \to \pi^0 N a)$ in \texttt{Madgraph}, we found that the flux of axions was approximately an order of magnitude less than that of $\pi^\pm N$-scattering in Fig.~\ref{fig:axionFlux}. It appears that this is due to the hierarchy of the low-energy coupling constants $C_{ap}$ and $C_{an}$; we can see this feature in the simpler $2 \to 2$ cross sections $\pi^0 n \to a n$, $\pi^0 p \to a p$, $\pi^+ n \to a p$, $\pi^+ p \to a n$.

In the gluon dominance model, $C_{ap} = -0.49$ and $C_{an} = -0.03$. To isolate the dependence on the proton and neutron couplings, we compare $\sigma(\pi^0 p \to a p)$, which only depends on $C_{ap}$,  to $\sigma(\pi^- p \to a n)$, which depends on both $C_{ap}$ and $C_{an}$. These two processes have similar Feynman diagrams, apart from an additional contact diagram for the $\pi^- p$ scattering. Since $C_{a \pi N} = 0$ when $C_{ap} = C_{an}$, we expect $\sigma(\pi^- p \to a n) = 2 \times \sigma(\pi^0 p \to a p)$ when the couplings are the same. Away from this limit, any large variation between the charged and neutral pion cross sections must be due to interference effects.

In Figure~\ref{fig:varyCan_varyEpi}, we plot the cross sections $\sigma(\pi^- p \to a n)$ and $\sigma(\pi^0 p \to a p)$ for a fixed $C_{ap} = -0.49$ while varying $C_{an} \in [-1,1]$. We chose to vary $C_{an}$ values because the neutral pion scattering does not depend on $C_{an}$. We see the expected behavior as we approach $C_{ap} = C_{an}$, but away from this limit, $\sigma(\pi^- p \to a n)$ can be orders of magnitude larger than $\sigma(\pi^0 p \to a p)$. At the physical value $C_{an} = -0.03$, the ratio of cross sections is \linebreak $\sigma(\pi^- p \to a n) / \sigma(\pi^0 p \to a p)$ = 2. We therefore conclude that the large hierarchy which we also observe in $2 \to 3$ scattering is likely due to interference effects.

\bibliography{references}

\begin{thebibliography}{51}%
\makeatletter
\providecommand \@ifxundefined [1]{%
 \@ifx{#1\undefined}
}%
\providecommand \@ifnum [1]{%
 \ifnum #1\expandafter \@firstoftwo
 \else \expandafter \@secondoftwo
 \fi
}%
\providecommand \@ifx [1]{%
 \ifx #1\expandafter \@firstoftwo
 \else \expandafter \@secondoftwo
 \fi
}%
\providecommand \natexlab [1]{#1}%
\providecommand \enquote  [1]{``#1''}%
\providecommand \bibnamefont  [1]{#1}%
\providecommand \bibfnamefont [1]{#1}%
\providecommand \citenamefont [1]{#1}%
\providecommand \href@noop [0]{\@secondoftwo}%
\providecommand \href [0]{\begingroup \@sanitize@url \@href}%
\providecommand \@href[1]{\@@startlink{#1}\@@href}%
\providecommand \@@href[1]{\endgroup#1\@@endlink}%
\providecommand \@sanitize@url [0]{\catcode `\\12\catcode `\$12\catcode
  `\&12\catcode `\#12\catcode `\^12\catcode `\_12\catcode `\%12\relax}%
\providecommand \@@startlink[1]{}%
\providecommand \@@endlink[0]{}%
\providecommand \url  [0]{\begingroup\@sanitize@url \@url }%
\providecommand \@url [1]{\endgroup\@href {#1}{\urlprefix }}%
\providecommand \urlprefix  [0]{URL }%
\providecommand \Eprint [0]{\href }%
\providecommand \doibase [0]{http://dx.doi.org/}%
\providecommand \selectlanguage [0]{\@gobble}%
\providecommand \bibinfo  [0]{\@secondoftwo}%
\providecommand \bibfield  [0]{\@secondoftwo}%
\providecommand \translation [1]{[#1]}%
\providecommand \BibitemOpen [0]{}%
\providecommand \bibitemStop [0]{}%
\providecommand \bibitemNoStop [0]{.\EOS\space}%
\providecommand \EOS [0]{\spacefactor3000\relax}%
\providecommand \BibitemShut  [1]{\csname bibitem#1\endcsname}%
\let\auto@bib@innerbib\@empty
\bibitem [{\citenamefont {Gori}\ \emph {et~al.}(2022)\citenamefont {Gori} \emph
  {et~al.}}]{Gori:2022vri}%
  \BibitemOpen
  \bibfield  {author} {\bibinfo {author} {\bibfnamefont {Stefania}\
  \bibnamefont {Gori}} \emph {et~al.},\ }\bibfield  {title} {\enquote {\bibinfo
  {title} {{Dark Sector Physics at High-Intensity Experiments}},}\ }\href@noop
  {} {\  (\bibinfo {year} {2022})},\ \Eprint {http://arxiv.org/abs/2209.04671}
  {arXiv:2209.04671 [hep-ph]} \BibitemShut {NoStop}%
\bibitem [{\citenamefont {Kelly}\ and\ \citenamefont
  {Tsai}(2019)}]{Kelly:2018brz}%
  \BibitemOpen
  \bibfield  {author} {\bibinfo {author} {\bibfnamefont {Kevin~J.}\
  \bibnamefont {Kelly}}\ and\ \bibinfo {author} {\bibfnamefont {Yu-Dai}\
  \bibnamefont {Tsai}},\ }\bibfield  {title} {\enquote {\bibinfo {title}
  {{Proton fixed-target scintillation experiment to search for millicharged
  dark matter}},}\ }\href {\doibase 10.1103/PhysRevD.100.015043} {\bibfield
  {journal} {\bibinfo  {journal} {Phys. Rev. D}\ }\textbf {\bibinfo {volume}
  {100}},\ \bibinfo {pages} {015043} (\bibinfo {year} {2019})},\ \Eprint
  {http://arxiv.org/abs/1812.03998} {arXiv:1812.03998 [hep-ph]} \BibitemShut
  {NoStop}%
\bibitem [{\citenamefont {Harnik}\ \emph {et~al.}(2019)\citenamefont {Harnik},
  \citenamefont {Liu},\ and\ \citenamefont {Palamara}}]{Harnik:2019zee}%
  \BibitemOpen
  \bibfield  {author} {\bibinfo {author} {\bibfnamefont {Roni}\ \bibnamefont
  {Harnik}}, \bibinfo {author} {\bibfnamefont {Zhen}\ \bibnamefont {Liu}}, \
  and\ \bibinfo {author} {\bibfnamefont {Ornella}\ \bibnamefont {Palamara}},\
  }\bibfield  {title} {\enquote {\bibinfo {title} {{Millicharged Particles in
  Liquid Argon Neutrino Experiments}},}\ }\href {\doibase
  10.1007/JHEP07(2019)170} {\bibfield  {journal} {\bibinfo  {journal} {JHEP}\
  }\textbf {\bibinfo {volume} {07}},\ \bibinfo {pages} {170} (\bibinfo {year}
  {2019})},\ \Eprint {http://arxiv.org/abs/1902.03246} {arXiv:1902.03246
  [hep-ph]} \BibitemShut {NoStop}%
\bibitem [{\citenamefont {Kelly}\ \emph {et~al.}(2021)\citenamefont {Kelly},
  \citenamefont {Kumar},\ and\ \citenamefont {Liu}}]{Kelly:2020dda}%
  \BibitemOpen
  \bibfield  {author} {\bibinfo {author} {\bibfnamefont {Kevin~J.}\
  \bibnamefont {Kelly}}, \bibinfo {author} {\bibfnamefont {Soubhik}\
  \bibnamefont {Kumar}}, \ and\ \bibinfo {author} {\bibfnamefont {Zhen}\
  \bibnamefont {Liu}},\ }\bibfield  {title} {\enquote {\bibinfo {title} {{Heavy
  axion opportunities at the DUNE near detector}},}\ }\href {\doibase
  10.1103/PhysRevD.103.095002} {\bibfield  {journal} {\bibinfo  {journal}
  {Phys. Rev. D}\ }\textbf {\bibinfo {volume} {103}},\ \bibinfo {pages}
  {095002} (\bibinfo {year} {2021})},\ \Eprint
  {http://arxiv.org/abs/2011.05995} {arXiv:2011.05995 [hep-ph]} \BibitemShut
  {NoStop}%
\bibitem [{\citenamefont {Berger}\ and\ \citenamefont
  {Putnam}(2024)}]{Berger:2024xqk}%
  \BibitemOpen
  \bibfield  {author} {\bibinfo {author} {\bibfnamefont {Joshua}\ \bibnamefont
  {Berger}}\ and\ \bibinfo {author} {\bibfnamefont {Gray}\ \bibnamefont
  {Putnam}},\ }\bibfield  {title} {\enquote {\bibinfo {title} {{Sensitivity to
  Kaon Decays to ALPs at Fixed Target Experiments}},}\ }\href@noop {} {\
  (\bibinfo {year} {2024})},\ \Eprint {http://arxiv.org/abs/2405.18480}
  {arXiv:2405.18480 [hep-ph]} \BibitemShut {NoStop}%
\bibitem [{\citenamefont {Acciarri}\ \emph {et~al.}(2015)\citenamefont
  {Acciarri} \emph {et~al.}}]{DUNE:2015lol}%
  \BibitemOpen
  \bibfield  {author} {\bibinfo {author} {\bibfnamefont {R.}~\bibnamefont
  {Acciarri}} \emph {et~al.} (\bibinfo {collaboration} {DUNE}),\ }\bibfield
  {title} {\enquote {\bibinfo {title} {{Long-Baseline Neutrino Facility (LBNF)
  and Deep Underground Neutrino Experiment (DUNE)}: {Conceptual Design Report,
  Volume 2: The Physics Program for DUNE at LBNF}},}\ }\href@noop {} {\
  (\bibinfo {year} {2015})},\ \Eprint {http://arxiv.org/abs/1512.06148}
  {arXiv:1512.06148 [physics.ins-det]} \BibitemShut {NoStop}%
\bibitem [{\citenamefont {Strait}\ \emph {et~al.}(2016)\citenamefont {Strait}
  \emph {et~al.}}]{DUNE:2016evb}%
  \BibitemOpen
  \bibfield  {author} {\bibinfo {author} {\bibfnamefont {James}\ \bibnamefont
  {Strait}} \emph {et~al.} (\bibinfo {collaboration} {DUNE}),\ }\bibfield
  {title} {\enquote {\bibinfo {title} {{Long-Baseline Neutrino Facility (LBNF)
  and Deep Underground Neutrino Experiment (DUNE)}: {Conceptual Design Report,
  Volume 3: Long-Baseline Neutrino Facility for DUNE June 24, 2015}},}\
  }\href@noop {} {\  (\bibinfo {year} {2016})},\ \Eprint
  {http://arxiv.org/abs/1601.05823} {arXiv:1601.05823 [physics.ins-det]}
  \BibitemShut {NoStop}%
\bibitem [{\citenamefont {Acciarri}\ \emph
  {et~al.}(2016{\natexlab{a}})\citenamefont {Acciarri} \emph
  {et~al.}}]{DUNE:2016hlj}%
  \BibitemOpen
  \bibfield  {author} {\bibinfo {author} {\bibfnamefont {R.}~\bibnamefont
  {Acciarri}} \emph {et~al.} (\bibinfo {collaboration} {DUNE}),\ }\bibfield
  {title} {\enquote {\bibinfo {title} {{Long-Baseline Neutrino Facility (LBNF)
  and Deep Underground Neutrino Experiment (DUNE)}: {Conceptual Design Report,
  Volume 1: The LBNF and DUNE Projects}},}\ }\href@noop {} {\  (\bibinfo {year}
  {2016}{\natexlab{a}})},\ \Eprint {http://arxiv.org/abs/1601.05471}
  {arXiv:1601.05471 [physics.ins-det]} \BibitemShut {NoStop}%
\bibitem [{\citenamefont {Acciarri}\ \emph
  {et~al.}(2016{\natexlab{b}})\citenamefont {Acciarri} \emph
  {et~al.}}]{DUNE:2016rla}%
  \BibitemOpen
  \bibfield  {author} {\bibinfo {author} {\bibfnamefont {R.}~\bibnamefont
  {Acciarri}} \emph {et~al.} (\bibinfo {collaboration} {DUNE}),\ }\bibfield
  {title} {\enquote {\bibinfo {title} {{Long-Baseline Neutrino Facility (LBNF)
  and Deep Underground Neutrino Experiment (DUNE)}: {Conceptual Design Report,
  Volume 4 The DUNE Detectors at LBNF}},}\ }\href@noop {} {\  (\bibinfo {year}
  {2016}{\natexlab{b}})},\ \Eprint {http://arxiv.org/abs/1601.02984}
  {arXiv:1601.02984 [physics.ins-det]} \BibitemShut {NoStop}%
\bibitem [{\citenamefont {Berlin}\ \emph {et~al.}(2018)\citenamefont {Berlin},
  \citenamefont {Gori}, \citenamefont {Schuster},\ and\ \citenamefont
  {Toro}}]{Berlin:2018pwi}%
  \BibitemOpen
  \bibfield  {author} {\bibinfo {author} {\bibfnamefont {Asher}\ \bibnamefont
  {Berlin}}, \bibinfo {author} {\bibfnamefont {Stefania}\ \bibnamefont {Gori}},
  \bibinfo {author} {\bibfnamefont {Philip}\ \bibnamefont {Schuster}}, \ and\
  \bibinfo {author} {\bibfnamefont {Natalia}\ \bibnamefont {Toro}},\ }\bibfield
   {title} {\enquote {\bibinfo {title} {{Dark Sectors at the Fermilab SeaQuest
  Experiment}},}\ }\href {\doibase 10.1103/PhysRevD.98.035011} {\bibfield
  {journal} {\bibinfo  {journal} {Phys. Rev. D}\ }\textbf {\bibinfo {volume}
  {98}},\ \bibinfo {pages} {035011} (\bibinfo {year} {2018})},\ \Eprint
  {http://arxiv.org/abs/1804.00661} {arXiv:1804.00661 [hep-ph]} \BibitemShut
  {NoStop}%
\bibitem [{\citenamefont {Curtin}\ \emph {et~al.}(2023)\citenamefont {Curtin},
  \citenamefont {Kahn},\ and\ \citenamefont {Nguyen}}]{Curtin:2023bcf}%
  \BibitemOpen
  \bibfield  {author} {\bibinfo {author} {\bibfnamefont {David}\ \bibnamefont
  {Curtin}}, \bibinfo {author} {\bibfnamefont {Yonatan}\ \bibnamefont {Kahn}},
  \ and\ \bibinfo {author} {\bibfnamefont {Rachel}\ \bibnamefont {Nguyen}},\
  }\bibfield  {title} {\enquote {\bibinfo {title} {{Dark photons from charged
  pion bremsstrahlung at proton beam experiments}},}\ }\href {\doibase
  10.1103/PhysRevD.108.095039} {\bibfield  {journal} {\bibinfo  {journal}
  {Phys. Rev. D}\ }\textbf {\bibinfo {volume} {108}},\ \bibinfo {pages}
  {095039} (\bibinfo {year} {2023})},\ \Eprint
  {http://arxiv.org/abs/2305.19309} {arXiv:2305.19309 [hep-ph]} \BibitemShut
  {NoStop}%
\bibitem [{\citenamefont {Berryman}\ \emph {et~al.}(2020)\citenamefont
  {Berryman}, \citenamefont {de~Gouvea}, \citenamefont {Fox}, \citenamefont
  {Kayser}, \citenamefont {Kelly},\ and\ \citenamefont
  {Raaf}}]{Berryman:2019dme}%
  \BibitemOpen
  \bibfield  {author} {\bibinfo {author} {\bibfnamefont {Jeffrey~M.}\
  \bibnamefont {Berryman}}, \bibinfo {author} {\bibfnamefont {Andre}\
  \bibnamefont {de~Gouvea}}, \bibinfo {author} {\bibfnamefont {Patrick~J}\
  \bibnamefont {Fox}}, \bibinfo {author} {\bibfnamefont {Boris~Jules}\
  \bibnamefont {Kayser}}, \bibinfo {author} {\bibfnamefont {Kevin~James}\
  \bibnamefont {Kelly}}, \ and\ \bibinfo {author} {\bibfnamefont
  {Jennifer~Lynne}\ \bibnamefont {Raaf}},\ }\bibfield  {title} {\enquote
  {\bibinfo {title} {{Searches for Decays of New Particles in the DUNE
  Multi-Purpose Near Detector}},}\ }\href {\doibase 10.1007/JHEP02(2020)174}
  {\bibfield  {journal} {\bibinfo  {journal} {JHEP}\ }\textbf {\bibinfo
  {volume} {02}},\ \bibinfo {pages} {174} (\bibinfo {year} {2020})},\ \Eprint
  {http://arxiv.org/abs/1912.07622} {arXiv:1912.07622 [hep-ph]} \BibitemShut
  {NoStop}%
\bibitem [{\citenamefont {Ruso}\ \emph {et~al.}(2022)\citenamefont {Ruso} \emph
  {et~al.}}]{Ruso:2022qes}%
  \BibitemOpen
  \bibfield  {author} {\bibinfo {author} {\bibfnamefont {L.~Alvarez}\
  \bibnamefont {Ruso}} \emph {et~al.},\ }\bibfield  {title} {\enquote {\bibinfo
  {title} {{Theoretical tools for neutrino scattering: interplay between
  lattice QCD, EFTs, nuclear physics, phenomenology, and neutrino event
  generators}},}\ }\href@noop {} {\  (\bibinfo {year} {2022})},\ \Eprint
  {http://arxiv.org/abs/2203.09030} {arXiv:2203.09030 [hep-ph]} \BibitemShut
  {NoStop}%
\bibitem [{\citenamefont {Ruiz}\ \emph {et~al.}(2024)\citenamefont {Ruiz} \emph
  {et~al.}}]{Ruiz:2023ozv}%
  \BibitemOpen
  \bibfield  {author} {\bibinfo {author} {\bibfnamefont {R.}~\bibnamefont
  {Ruiz}} \emph {et~al.},\ }\bibfield  {title} {\enquote {\bibinfo {title}
  {{Target mass corrections in lepton\textendash{}nucleus DIS: Theory and
  applications to nuclear PDFs}},}\ }\href {\doibase
  10.1016/j.ppnp.2023.104096} {\bibfield  {journal} {\bibinfo  {journal} {Prog.
  Part. Nucl. Phys.}\ }\textbf {\bibinfo {volume} {136}},\ \bibinfo {pages}
  {104096} (\bibinfo {year} {2024})},\ \Eprint
  {http://arxiv.org/abs/2301.07715} {arXiv:2301.07715 [hep-ph]} \BibitemShut
  {NoStop}%
\bibitem [{\citenamefont {Alwall}\ \emph {et~al.}(2014)\citenamefont {Alwall},
  \citenamefont {Frederix}, \citenamefont {Frixione}, \citenamefont {Hirschi},
  \citenamefont {Maltoni}, \citenamefont {Mattelaer}, \citenamefont {Shao},
  \citenamefont {Stelzer}, \citenamefont {Torrielli},\ and\ \citenamefont
  {Zaro}}]{Alwall:2014hca}%
  \BibitemOpen
  \bibfield  {author} {\bibinfo {author} {\bibfnamefont {J.}~\bibnamefont
  {Alwall}}, \bibinfo {author} {\bibfnamefont {R.}~\bibnamefont {Frederix}},
  \bibinfo {author} {\bibfnamefont {S.}~\bibnamefont {Frixione}}, \bibinfo
  {author} {\bibfnamefont {V.}~\bibnamefont {Hirschi}}, \bibinfo {author}
  {\bibfnamefont {F.}~\bibnamefont {Maltoni}}, \bibinfo {author} {\bibfnamefont
  {O.}~\bibnamefont {Mattelaer}}, \bibinfo {author} {\bibfnamefont {H.~S.}\
  \bibnamefont {Shao}}, \bibinfo {author} {\bibfnamefont {T.}~\bibnamefont
  {Stelzer}}, \bibinfo {author} {\bibfnamefont {P.}~\bibnamefont {Torrielli}},
  \ and\ \bibinfo {author} {\bibfnamefont {M.}~\bibnamefont {Zaro}},\
  }\bibfield  {title} {\enquote {\bibinfo {title} {{The automated computation
  of tree-level and next-to-leading order differential cross sections, and
  their matching to parton shower simulations}},}\ }\href {\doibase
  10.1007/JHEP07(2014)079} {\bibfield  {journal} {\bibinfo  {journal} {JHEP}\
  }\textbf {\bibinfo {volume} {07}},\ \bibinfo {pages} {079} (\bibinfo {year}
  {2014})},\ \Eprint {http://arxiv.org/abs/1405.0301} {arXiv:1405.0301
  [hep-ph]} \BibitemShut {NoStop}%
\bibitem [{\citenamefont {Scherer}(2003)}]{Scherer:2002tk}%
  \BibitemOpen
  \bibfield  {author} {\bibinfo {author} {\bibfnamefont {Stefan}\ \bibnamefont
  {Scherer}},\ }\bibfield  {title} {\enquote {\bibinfo {title} {{Introduction
  to chiral perturbation theory}},}\ }\href@noop {} {\bibfield  {journal}
  {\bibinfo  {journal} {Adv. Nucl. Phys.}\ }\textbf {\bibinfo {volume} {27}},\
  \bibinfo {pages} {277} (\bibinfo {year} {2003})},\ \Eprint
  {http://arxiv.org/abs/hep-ph/0210398} {arXiv:hep-ph/0210398} \BibitemShut
  {NoStop}%
\bibitem [{\citenamefont {Shin}\ and\ \citenamefont
  {Yun}(2022)}]{Shin:2022ulh}%
  \BibitemOpen
  \bibfield  {author} {\bibinfo {author} {\bibfnamefont {Chang~Sub}\
  \bibnamefont {Shin}}\ and\ \bibinfo {author} {\bibfnamefont {Seokhoon}\
  \bibnamefont {Yun}},\ }\bibfield  {title} {\enquote {\bibinfo {title} {{Dark
  gauge boson emission from supernova pions}},}\ }\href@noop {} {\  (\bibinfo
  {year} {2022})},\ \Eprint {http://arxiv.org/abs/2211.15677} {arXiv:2211.15677
  [hep-ph]} \BibitemShut {NoStop}%
\bibitem [{\citenamefont {Ellis}\ and\ \citenamefont
  {Tang}(1998)}]{Ellis:1997kc}%
  \BibitemOpen
  \bibfield  {author} {\bibinfo {author} {\bibfnamefont {Paul~J.}\ \bibnamefont
  {Ellis}}\ and\ \bibinfo {author} {\bibfnamefont {Hua-Bin}\ \bibnamefont
  {Tang}},\ }\bibfield  {title} {\enquote {\bibinfo {title} {{Pion nucleon
  scattering in a new approach to chiral perturbation theory}},}\ }\href
  {\doibase 10.1103/PhysRevC.57.3356} {\bibfield  {journal} {\bibinfo
  {journal} {Phys. Rev. C}\ }\textbf {\bibinfo {volume} {57}},\ \bibinfo
  {pages} {3356--3375} (\bibinfo {year} {1998})},\ \Eprint
  {http://arxiv.org/abs/hep-ph/9709354} {arXiv:hep-ph/9709354} \BibitemShut
  {NoStop}%
\bibitem [{\citenamefont {Bernard}\ \emph {et~al.}(1995)\citenamefont
  {Bernard}, \citenamefont {Kaiser},\ and\ \citenamefont
  {Meissner}}]{Bernard:1995gx}%
  \BibitemOpen
  \bibfield  {author} {\bibinfo {author} {\bibfnamefont {V.}~\bibnamefont
  {Bernard}}, \bibinfo {author} {\bibfnamefont {Norbert}\ \bibnamefont
  {Kaiser}}, \ and\ \bibinfo {author} {\bibfnamefont {Ulf~G.}\ \bibnamefont
  {Meissner}},\ }\bibfield  {title} {\enquote {\bibinfo {title} {{The Reaction
  $\pi N \to \pi \pi N$ at threshold in chiral perturbation theory}},}\ }\href
  {\doibase 10.1016/0550-3213(95)00526-9} {\bibfield  {journal} {\bibinfo
  {journal} {Nucl. Phys. B}\ }\textbf {\bibinfo {volume} {457}},\ \bibinfo
  {pages} {147--172} (\bibinfo {year} {1995})},\ \Eprint
  {http://arxiv.org/abs/hep-ph/9507418} {arXiv:hep-ph/9507418} \BibitemShut
  {NoStop}%
\bibitem [{\citenamefont {Bernard}\ \emph {et~al.}(1997)\citenamefont
  {Bernard}, \citenamefont {Kaiser},\ and\ \citenamefont
  {Meissner}}]{Bernard:1997tq}%
  \BibitemOpen
  \bibfield  {author} {\bibinfo {author} {\bibfnamefont {V.}~\bibnamefont
  {Bernard}}, \bibinfo {author} {\bibfnamefont {Norbert}\ \bibnamefont
  {Kaiser}}, \ and\ \bibinfo {author} {\bibfnamefont {Ulf-G.}\ \bibnamefont
  {Meissner}},\ }\bibfield  {title} {\enquote {\bibinfo {title} {{The Reaction
  $\pi N \to \pi \pi N$ above threshold in chiral perturbation theory}},}\
  }\href {\doibase 10.1016/S0375-9474(97)00159-0} {\bibfield  {journal}
  {\bibinfo  {journal} {Nucl. Phys. A}\ }\textbf {\bibinfo {volume} {619}},\
  \bibinfo {pages} {261--284} (\bibinfo {year} {1997})},\ \Eprint
  {http://arxiv.org/abs/hep-ph/9703218} {arXiv:hep-ph/9703218} \BibitemShut
  {NoStop}%
\bibitem [{\citenamefont {Czarnecki}\ \emph {et~al.}(2018)\citenamefont
  {Czarnecki}, \citenamefont {Marciano},\ and\ \citenamefont
  {Sirlin}}]{Czarnecki:2018okw}%
  \BibitemOpen
  \bibfield  {author} {\bibinfo {author} {\bibfnamefont {Andrzej}\ \bibnamefont
  {Czarnecki}}, \bibinfo {author} {\bibfnamefont {William~J.}\ \bibnamefont
  {Marciano}}, \ and\ \bibinfo {author} {\bibfnamefont {Alberto}\ \bibnamefont
  {Sirlin}},\ }\bibfield  {title} {\enquote {\bibinfo {title} {{Neutron
  Lifetime and Axial Coupling Connection}},}\ }\href {\doibase
  10.1103/PhysRevLett.120.202002} {\bibfield  {journal} {\bibinfo  {journal}
  {Phys. Rev. Lett.}\ }\textbf {\bibinfo {volume} {120}},\ \bibinfo {pages}
  {202002} (\bibinfo {year} {2018})},\ \Eprint
  {http://arxiv.org/abs/1802.01804} {arXiv:1802.01804 [hep-ph]} \BibitemShut
  {NoStop}%
\bibitem [{\citenamefont {Haas}\ \emph {et~al.}(2015)\citenamefont {Haas},
  \citenamefont {Hill}, \citenamefont {Izaguirre},\ and\ \citenamefont
  {Yavin}}]{Haas:2014dda}%
  \BibitemOpen
  \bibfield  {author} {\bibinfo {author} {\bibfnamefont {Andrew}\ \bibnamefont
  {Haas}}, \bibinfo {author} {\bibfnamefont {Christopher~S.}\ \bibnamefont
  {Hill}}, \bibinfo {author} {\bibfnamefont {Eder}\ \bibnamefont {Izaguirre}},
  \ and\ \bibinfo {author} {\bibfnamefont {Itay}\ \bibnamefont {Yavin}},\
  }\bibfield  {title} {\enquote {\bibinfo {title} {{Looking for milli-charged
  particles with a new experiment at the LHC}},}\ }\href {\doibase
  10.1016/j.physletb.2015.04.062} {\bibfield  {journal} {\bibinfo  {journal}
  {Phys. Lett. B}\ }\textbf {\bibinfo {volume} {746}},\ \bibinfo {pages}
  {117--120} (\bibinfo {year} {2015})},\ \Eprint
  {http://arxiv.org/abs/1410.6816} {arXiv:1410.6816 [hep-ph]} \BibitemShut
  {NoStop}%
\bibitem [{\citenamefont {Ball}\ \emph {et~al.}(2016)\citenamefont {Ball} \emph
  {et~al.}}]{Ball:2016zrp}%
  \BibitemOpen
  \bibfield  {author} {\bibinfo {author} {\bibfnamefont {Austin}\ \bibnamefont
  {Ball}} \emph {et~al.},\ }\bibfield  {title} {\enquote {\bibinfo {title} {{A
  Letter of Intent to Install a milli-charged Particle Detector at LHC P5}},}\
  }\href@noop {} {\  (\bibinfo {year} {2016})},\ \Eprint
  {http://arxiv.org/abs/1607.04669} {arXiv:1607.04669 [physics.ins-det]}
  \BibitemShut {NoStop}%
\bibitem [{\citenamefont {Patrignani}\ \emph {et~al.}(2016)\citenamefont
  {Patrignani} \emph {et~al.}}]{ParticleDataGroup:2016lqr}%
  \BibitemOpen
  \bibfield  {author} {\bibinfo {author} {\bibfnamefont {C.}~\bibnamefont
  {Patrignani}} \emph {et~al.} (\bibinfo {collaboration} {Particle Data
  Group}),\ }\bibfield  {title} {\enquote {\bibinfo {title} {{Review of
  Particle Physics}},}\ }\href {\doibase 10.1088/1674-1137/40/10/100001}
  {\bibfield  {journal} {\bibinfo  {journal} {Chin. Phys. C}\ }\textbf
  {\bibinfo {volume} {40}},\ \bibinfo {pages} {100001} (\bibinfo {year}
  {2016})}\BibitemShut {NoStop}%
\bibitem [{\citenamefont {Kelly}()}]{KellyPrivate}%
  \BibitemOpen
  \bibfield  {author} {\bibinfo {author} {\bibfnamefont {Kevin}\ \bibnamefont
  {Kelly}},\ }\href@noop {} {}\bibinfo {howpublished} {private
  communication}\BibitemShut {NoStop}%
\bibitem [{\citenamefont {Tsai}\ \emph {et~al.}(2024)\citenamefont {Tsai},
  \citenamefont {Hwang}, \citenamefont {Schmitz}, \citenamefont {Citron},
  \citenamefont {Gunthoti}, \citenamefont {Steenis}, \citenamefont {Jeong},
  \citenamefont {Moon}, \citenamefont {Yoo},\ and\ \citenamefont
  {Liu}}]{Tsai:2024wdh}%
  \BibitemOpen
  \bibfield  {author} {\bibinfo {author} {\bibfnamefont {Yu-Dai}\ \bibnamefont
  {Tsai}}, \bibinfo {author} {\bibfnamefont {Insung}\ \bibnamefont {Hwang}},
  \bibinfo {author} {\bibfnamefont {Ryan}\ \bibnamefont {Schmitz}}, \bibinfo
  {author} {\bibfnamefont {Matthew}\ \bibnamefont {Citron}}, \bibinfo {author}
  {\bibfnamefont {Kranti}\ \bibnamefont {Gunthoti}}, \bibinfo {author}
  {\bibfnamefont {Jacob}\ \bibnamefont {Steenis}}, \bibinfo {author}
  {\bibfnamefont {Hoyong}\ \bibnamefont {Jeong}}, \bibinfo {author}
  {\bibfnamefont {Hyunki}\ \bibnamefont {Moon}}, \bibinfo {author}
  {\bibfnamefont {Jae~Hyeok}\ \bibnamefont {Yoo}}, \ and\ \bibinfo {author}
  {\bibfnamefont {Ming~Xiong}\ \bibnamefont {Liu}},\ }\bibfield  {title}
  {\enquote {\bibinfo {title} {{LANSCE-mQ: Dedicated search for
  milli/fractionally charged particles at LANL}},}\ }\href@noop {} {\
  (\bibinfo {year} {2024})},\ \Eprint {http://arxiv.org/abs/2407.07142}
  {arXiv:2407.07142 [hep-ph]} \BibitemShut {NoStop}%
\bibitem [{\citenamefont {Prinz}(2001)}]{Prinz:2001qz}%
  \BibitemOpen
  \bibfield  {author} {\bibinfo {author} {\bibfnamefont {Alyssa~Ann}\
  \bibnamefont {Prinz}},\ }\emph {\bibinfo {title} {{The Search for
  millicharged particles at SLAC}}},\ \href@noop {} {\bibinfo {type} {Other
  thesis}} (\bibinfo {year} {2001})\BibitemShut {NoStop}%
\bibitem [{\citenamefont {Abramoff}\ \emph {et~al.}(2019)\citenamefont
  {Abramoff} \emph {et~al.}}]{SENSEI:2019ibb}%
  \BibitemOpen
  \bibfield  {author} {\bibinfo {author} {\bibfnamefont {Orr}\ \bibnamefont
  {Abramoff}} \emph {et~al.} (\bibinfo {collaboration} {SENSEI}),\ }\bibfield
  {title} {\enquote {\bibinfo {title} {{SENSEI: Direct-Detection Constraints on
  Sub-GeV Dark Matter from a Shallow Underground Run Using a Prototype
  Skipper-CCD}},}\ }\href {\doibase 10.1103/PhysRevLett.122.161801} {\bibfield
  {journal} {\bibinfo  {journal} {Phys. Rev. Lett.}\ }\textbf {\bibinfo
  {volume} {122}},\ \bibinfo {pages} {161801} (\bibinfo {year} {2019})},\
  \Eprint {http://arxiv.org/abs/1901.10478} {arXiv:1901.10478 [hep-ex]}
  \BibitemShut {NoStop}%
\bibitem [{\citenamefont {Marocco}\ and\ \citenamefont
  {Sarkar}(2021)}]{Marocco:2020dqu}%
  \BibitemOpen
  \bibfield  {author} {\bibinfo {author} {\bibfnamefont {Giacomo}\ \bibnamefont
  {Marocco}}\ and\ \bibinfo {author} {\bibfnamefont {Subir}\ \bibnamefont
  {Sarkar}},\ }\bibfield  {title} {\enquote {\bibinfo {title} {{Blast from the
  past: Constraints on the dark sector from the BEBC WA66 beam dump
  experiment}},}\ }\href {\doibase 10.21468/SciPostPhys.10.2.043} {\bibfield
  {journal} {\bibinfo  {journal} {SciPost Phys.}\ }\textbf {\bibinfo {volume}
  {10}},\ \bibinfo {pages} {043} (\bibinfo {year} {2021})},\ \Eprint
  {http://arxiv.org/abs/2011.08153} {arXiv:2011.08153 [hep-ph]} \BibitemShut
  {NoStop}%
\bibitem [{\citenamefont {De~Winter}\ \emph {et~al.}(1989)\citenamefont
  {De~Winter} \emph {et~al.}}]{CHARM-II:1989nic}%
  \BibitemOpen
  \bibfield  {author} {\bibinfo {author} {\bibfnamefont {K.}~\bibnamefont
  {De~Winter}} \emph {et~al.} (\bibinfo {collaboration} {CHARM-II}),\
  }\bibfield  {title} {\enquote {\bibinfo {title} {{A Detector for the Study of
  Neutrino - Electron Scattering}},}\ }\href {\doibase
  10.1016/0168-9002(89)91190-X} {\bibfield  {journal} {\bibinfo  {journal}
  {Nucl. Instrum. Meth. A}\ }\textbf {\bibinfo {volume} {278}},\ \bibinfo
  {pages} {670} (\bibinfo {year} {1989})}\BibitemShut {NoStop}%
\bibitem [{\citenamefont {Magill}\ \emph {et~al.}(2019)\citenamefont {Magill},
  \citenamefont {Plestid}, \citenamefont {Pospelov},\ and\ \citenamefont
  {Tsai}}]{Magill:2018tbb}%
  \BibitemOpen
  \bibfield  {author} {\bibinfo {author} {\bibfnamefont {Gabriel}\ \bibnamefont
  {Magill}}, \bibinfo {author} {\bibfnamefont {Ryan}\ \bibnamefont {Plestid}},
  \bibinfo {author} {\bibfnamefont {Maxim}\ \bibnamefont {Pospelov}}, \ and\
  \bibinfo {author} {\bibfnamefont {Yu-Dai}\ \bibnamefont {Tsai}},\ }\bibfield
  {title} {\enquote {\bibinfo {title} {{Millicharged particles in neutrino
  experiments}},}\ }\href {\doibase 10.1103/PhysRevLett.122.071801} {\bibfield
  {journal} {\bibinfo  {journal} {Phys. Rev. Lett.}\ }\textbf {\bibinfo
  {volume} {122}},\ \bibinfo {pages} {071801} (\bibinfo {year} {2019})},\
  \Eprint {http://arxiv.org/abs/1806.03310} {arXiv:1806.03310 [hep-ph]}
  \BibitemShut {NoStop}%
\bibitem [{\citenamefont {Acciarri}\ \emph {et~al.}(2020)\citenamefont
  {Acciarri} \emph {et~al.}}]{ArgoNeuT:2019ckq}%
  \BibitemOpen
  \bibfield  {author} {\bibinfo {author} {\bibfnamefont {R.}~\bibnamefont
  {Acciarri}} \emph {et~al.} (\bibinfo {collaboration} {ArgoNeuT}),\ }\bibfield
   {title} {\enquote {\bibinfo {title} {{Improved Limits on Millicharged
  Particles Using the ArgoNeuT Experiment at Fermilab}},}\ }\href {\doibase
  10.1103/PhysRevLett.124.131801} {\bibfield  {journal} {\bibinfo  {journal}
  {Phys. Rev. Lett.}\ }\textbf {\bibinfo {volume} {124}},\ \bibinfo {pages}
  {131801} (\bibinfo {year} {2020})},\ \Eprint
  {http://arxiv.org/abs/1911.07996} {arXiv:1911.07996 [hep-ex]} \BibitemShut
  {NoStop}%
\bibitem [{\citenamefont {Davidson}\ \emph {et~al.}(2000)\citenamefont
  {Davidson}, \citenamefont {Hannestad},\ and\ \citenamefont
  {Raffelt}}]{Davidson:2000hf}%
  \BibitemOpen
  \bibfield  {author} {\bibinfo {author} {\bibfnamefont {Sacha}\ \bibnamefont
  {Davidson}}, \bibinfo {author} {\bibfnamefont {Steen}\ \bibnamefont
  {Hannestad}}, \ and\ \bibinfo {author} {\bibfnamefont {Georg}\ \bibnamefont
  {Raffelt}},\ }\bibfield  {title} {\enquote {\bibinfo {title} {{Updated bounds
  on millicharged particles}},}\ }\href {\doibase
  10.1088/1126-6708/2000/05/003} {\bibfield  {journal} {\bibinfo  {journal}
  {JHEP}\ }\textbf {\bibinfo {volume} {05}},\ \bibinfo {pages} {003} (\bibinfo
  {year} {2000})},\ \Eprint {http://arxiv.org/abs/hep-ph/0001179}
  {arXiv:hep-ph/0001179} \BibitemShut {NoStop}%
\bibitem [{\citenamefont {Hook}\ \emph {et~al.}(2020)\citenamefont {Hook},
  \citenamefont {Kumar}, \citenamefont {Liu},\ and\ \citenamefont
  {Sundrum}}]{Hook:2019qoh}%
  \BibitemOpen
  \bibfield  {author} {\bibinfo {author} {\bibfnamefont {Anson}\ \bibnamefont
  {Hook}}, \bibinfo {author} {\bibfnamefont {Soubhik}\ \bibnamefont {Kumar}},
  \bibinfo {author} {\bibfnamefont {Zhen}\ \bibnamefont {Liu}}, \ and\ \bibinfo
  {author} {\bibfnamefont {Raman}\ \bibnamefont {Sundrum}},\ }\bibfield
  {title} {\enquote {\bibinfo {title} {{High Quality QCD Axion and the LHC}},}\
  }\href {\doibase 10.1103/PhysRevLett.124.221801} {\bibfield  {journal}
  {\bibinfo  {journal} {Phys. Rev. Lett.}\ }\textbf {\bibinfo {volume} {124}},\
  \bibinfo {pages} {221801} (\bibinfo {year} {2020})},\ \Eprint
  {http://arxiv.org/abs/1911.12364} {arXiv:1911.12364 [hep-ph]} \BibitemShut
  {NoStop}%
\bibitem [{\citenamefont {Di~Luzio}\ \emph {et~al.}(2020)\citenamefont
  {Di~Luzio}, \citenamefont {Giannotti}, \citenamefont {Nardi},\ and\
  \citenamefont {Visinelli}}]{DiLuzio:2020wdo}%
  \BibitemOpen
  \bibfield  {author} {\bibinfo {author} {\bibfnamefont {Luca}\ \bibnamefont
  {Di~Luzio}}, \bibinfo {author} {\bibfnamefont {Maurizio}\ \bibnamefont
  {Giannotti}}, \bibinfo {author} {\bibfnamefont {Enrico}\ \bibnamefont
  {Nardi}}, \ and\ \bibinfo {author} {\bibfnamefont {Luca}\ \bibnamefont
  {Visinelli}},\ }\bibfield  {title} {\enquote {\bibinfo {title} {{The
  landscape of QCD axion models}},}\ }\href {\doibase
  10.1016/j.physrep.2020.06.002} {\bibfield  {journal} {\bibinfo  {journal}
  {Phys. Rept.}\ }\textbf {\bibinfo {volume} {870}},\ \bibinfo {pages} {1--117}
  (\bibinfo {year} {2020})},\ \Eprint {http://arxiv.org/abs/2003.01100}
  {arXiv:2003.01100 [hep-ph]} \BibitemShut {NoStop}%
\bibitem [{\citenamefont {Chang}\ and\ \citenamefont
  {Choi}(1993)}]{Chang:1993gm}%
  \BibitemOpen
  \bibfield  {author} {\bibinfo {author} {\bibfnamefont {Sanghyeon}\
  \bibnamefont {Chang}}\ and\ \bibinfo {author} {\bibfnamefont {Kiwoon}\
  \bibnamefont {Choi}},\ }\bibfield  {title} {\enquote {\bibinfo {title}
  {{Hadronic axion window and the big bang nucleosynthesis}},}\ }\href
  {\doibase 10.1016/0370-2693(93)90656-3} {\bibfield  {journal} {\bibinfo
  {journal} {Phys. Lett. B}\ }\textbf {\bibinfo {volume} {316}},\ \bibinfo
  {pages} {51--56} (\bibinfo {year} {1993})},\ \Eprint
  {http://arxiv.org/abs/hep-ph/9306216} {arXiv:hep-ph/9306216} \BibitemShut
  {NoStop}%
\bibitem [{\citenamefont {Choi}\ \emph {et~al.}(2022)\citenamefont {Choi},
  \citenamefont {Kim}, \citenamefont {Seong},\ and\ \citenamefont
  {Shin}}]{Choi:2021ign}%
  \BibitemOpen
  \bibfield  {author} {\bibinfo {author} {\bibfnamefont {Kiwoon}\ \bibnamefont
  {Choi}}, \bibinfo {author} {\bibfnamefont {Hee~Jung}\ \bibnamefont {Kim}},
  \bibinfo {author} {\bibfnamefont {Hyeonseok}\ \bibnamefont {Seong}}, \ and\
  \bibinfo {author} {\bibfnamefont {Chang~Sub}\ \bibnamefont {Shin}},\
  }\bibfield  {title} {\enquote {\bibinfo {title} {{Axion emission from
  supernova with axion-pion-nucleon contact interaction}},}\ }\href {\doibase
  10.1007/JHEP02(2022)143} {\bibfield  {journal} {\bibinfo  {journal} {JHEP}\
  }\textbf {\bibinfo {volume} {02}},\ \bibinfo {pages} {143} (\bibinfo {year}
  {2022})},\ \Eprint {http://arxiv.org/abs/2110.01972} {arXiv:2110.01972
  [hep-ph]} \BibitemShut {NoStop}%
\bibitem [{\citenamefont {Grilli~di Cortona}\ \emph {et~al.}(2016)\citenamefont
  {Grilli~di Cortona}, \citenamefont {Hardy}, \citenamefont {Pardo~Vega},\ and\
  \citenamefont {Villadoro}}]{GrillidiCortona:2015jxo}%
  \BibitemOpen
  \bibfield  {author} {\bibinfo {author} {\bibfnamefont {Giovanni}\
  \bibnamefont {Grilli~di Cortona}}, \bibinfo {author} {\bibfnamefont {Edward}\
  \bibnamefont {Hardy}}, \bibinfo {author} {\bibfnamefont {Javier}\
  \bibnamefont {Pardo~Vega}}, \ and\ \bibinfo {author} {\bibfnamefont
  {Giovanni}\ \bibnamefont {Villadoro}},\ }\bibfield  {title} {\enquote
  {\bibinfo {title} {{The QCD axion, precisely}},}\ }\href {\doibase
  10.1007/JHEP01(2016)034} {\bibfield  {journal} {\bibinfo  {journal} {JHEP}\
  }\textbf {\bibinfo {volume} {01}},\ \bibinfo {pages} {034} (\bibinfo {year}
  {2016})},\ \Eprint {http://arxiv.org/abs/1511.02867} {arXiv:1511.02867
  [hep-ph]} \BibitemShut {NoStop}%
\bibitem [{\citenamefont {Aloni}\ \emph {et~al.}(2019)\citenamefont {Aloni},
  \citenamefont {Soreq},\ and\ \citenamefont {Williams}}]{Aloni:2018vki}%
  \BibitemOpen
  \bibfield  {author} {\bibinfo {author} {\bibfnamefont {Daniel}\ \bibnamefont
  {Aloni}}, \bibinfo {author} {\bibfnamefont {Yotam}\ \bibnamefont {Soreq}}, \
  and\ \bibinfo {author} {\bibfnamefont {Mike}\ \bibnamefont {Williams}},\
  }\bibfield  {title} {\enquote {\bibinfo {title} {{Coupling QCD-Scale
  Axionlike Particles to Gluons}},}\ }\href {\doibase
  10.1103/PhysRevLett.123.031803} {\bibfield  {journal} {\bibinfo  {journal}
  {Phys. Rev. Lett.}\ }\textbf {\bibinfo {volume} {123}},\ \bibinfo {pages}
  {031803} (\bibinfo {year} {2019})},\ \Eprint
  {http://arxiv.org/abs/1811.03474} {arXiv:1811.03474 [hep-ph]} \BibitemShut
  {NoStop}%
\bibitem [{\citenamefont {Blinov}\ \emph {et~al.}(2022)\citenamefont {Blinov},
  \citenamefont {Kowalczyk},\ and\ \citenamefont {Wynne}}]{Blinov:2021say}%
  \BibitemOpen
  \bibfield  {author} {\bibinfo {author} {\bibfnamefont {Nikita}\ \bibnamefont
  {Blinov}}, \bibinfo {author} {\bibfnamefont {Elizabeth}\ \bibnamefont
  {Kowalczyk}}, \ and\ \bibinfo {author} {\bibfnamefont {Margaret}\
  \bibnamefont {Wynne}},\ }\bibfield  {title} {\enquote {\bibinfo {title}
  {{Axion-like particle searches at DarkQuest}},}\ }\href {\doibase
  10.1007/JHEP02(2022)036} {\bibfield  {journal} {\bibinfo  {journal} {JHEP}\
  }\textbf {\bibinfo {volume} {02}},\ \bibinfo {pages} {036} (\bibinfo {year}
  {2022})},\ \Eprint {http://arxiv.org/abs/2112.09814} {arXiv:2112.09814
  [hep-ph]} \BibitemShut {NoStop}%
\bibitem [{\citenamefont {Ariga}\ \emph {et~al.}(2019)\citenamefont {Ariga}
  \emph {et~al.}}]{FASER:2018eoc}%
  \BibitemOpen
  \bibfield  {author} {\bibinfo {author} {\bibfnamefont {Akitaka}\ \bibnamefont
  {Ariga}} \emph {et~al.} (\bibinfo {collaboration} {FASER}),\ }\bibfield
  {title} {\enquote {\bibinfo {title} {{FASER\textquoteright{}s physics reach
  for long-lived particles}},}\ }\href {\doibase 10.1103/PhysRevD.99.095011}
  {\bibfield  {journal} {\bibinfo  {journal} {Phys. Rev. D}\ }\textbf {\bibinfo
  {volume} {99}},\ \bibinfo {pages} {095011} (\bibinfo {year} {2019})},\
  \Eprint {http://arxiv.org/abs/1811.12522} {arXiv:1811.12522 [hep-ph]}
  \BibitemShut {NoStop}%
\bibitem [{\citenamefont {Cortina~Gil}\ \emph {et~al.}(2021)\citenamefont
  {Cortina~Gil} \emph {et~al.}}]{NA62:2021zjw}%
  \BibitemOpen
  \bibfield  {author} {\bibinfo {author} {\bibfnamefont {Eduardo}\ \bibnamefont
  {Cortina~Gil}} \emph {et~al.} (\bibinfo {collaboration} {NA62}),\ }\bibfield
  {title} {\enquote {\bibinfo {title} {{Measurement of the very rare
  K$^{+}$\textrightarrow{}$ {\pi}^{+}\nu \overline{\nu} $ decay}},}\ }\href
  {\doibase 10.1007/JHEP06(2021)093} {\bibfield  {journal} {\bibinfo  {journal}
  {JHEP}\ }\textbf {\bibinfo {volume} {06}},\ \bibinfo {pages} {093} (\bibinfo
  {year} {2021})},\ \Eprint {http://arxiv.org/abs/2103.15389} {arXiv:2103.15389
  [hep-ex]} \BibitemShut {NoStop}%
\bibitem [{\citenamefont {Ertas}\ and\ \citenamefont
  {Kahlhoefer}(2020)}]{Ertas:2020xcc}%
  \BibitemOpen
  \bibfield  {author} {\bibinfo {author} {\bibfnamefont {Fatih}\ \bibnamefont
  {Ertas}}\ and\ \bibinfo {author} {\bibfnamefont {Felix}\ \bibnamefont
  {Kahlhoefer}},\ }\bibfield  {title} {\enquote {\bibinfo {title} {{On the
  interplay between astrophysical and laboratory probes of MeV-scale axion-like
  particles}},}\ }\href {\doibase 10.1007/JHEP07(2020)050} {\bibfield
  {journal} {\bibinfo  {journal} {JHEP}\ }\textbf {\bibinfo {volume} {07}},\
  \bibinfo {pages} {050} (\bibinfo {year} {2020})},\ \Eprint
  {http://arxiv.org/abs/2004.01193} {arXiv:2004.01193 [hep-ph]} \BibitemShut
  {NoStop}%
\bibitem [{\citenamefont {Chang}\ \emph {et~al.}(2018)\citenamefont {Chang},
  \citenamefont {Essig},\ and\ \citenamefont {McDermott}}]{Chang:2018rso}%
  \BibitemOpen
  \bibfield  {author} {\bibinfo {author} {\bibfnamefont {Jae~Hyeok}\
  \bibnamefont {Chang}}, \bibinfo {author} {\bibfnamefont {Rouven}\
  \bibnamefont {Essig}}, \ and\ \bibinfo {author} {\bibfnamefont {Samuel~D.}\
  \bibnamefont {McDermott}},\ }\bibfield  {title} {\enquote {\bibinfo {title}
  {{Supernova 1987A Constraints on Sub-GeV Dark Sectors, Millicharged
  Particles, the QCD Axion, and an Axion-like Particle}},}\ }\href {\doibase
  10.1007/JHEP09(2018)051} {\bibfield  {journal} {\bibinfo  {journal} {JHEP}\
  }\textbf {\bibinfo {volume} {09}},\ \bibinfo {pages} {051} (\bibinfo {year}
  {2018})},\ \Eprint {http://arxiv.org/abs/1803.00993} {arXiv:1803.00993
  [hep-ph]} \BibitemShut {NoStop}%
\bibitem [{\citenamefont {Depta}\ \emph {et~al.}(2020)\citenamefont {Depta},
  \citenamefont {Hufnagel},\ and\ \citenamefont
  {Schmidt-Hoberg}}]{Depta:2020wmr}%
  \BibitemOpen
  \bibfield  {author} {\bibinfo {author} {\bibfnamefont {Paul~Frederik}\
  \bibnamefont {Depta}}, \bibinfo {author} {\bibfnamefont {Marco}\ \bibnamefont
  {Hufnagel}}, \ and\ \bibinfo {author} {\bibfnamefont {Kai}\ \bibnamefont
  {Schmidt-Hoberg}},\ }\bibfield  {title} {\enquote {\bibinfo {title} {{Robust
  cosmological constraints on axion-like particles}},}\ }\href {\doibase
  10.1088/1475-7516/2020/05/009} {\bibfield  {journal} {\bibinfo  {journal}
  {JCAP}\ }\textbf {\bibinfo {volume} {05}},\ \bibinfo {pages} {009} (\bibinfo
  {year} {2020})},\ \Eprint {http://arxiv.org/abs/2002.08370} {arXiv:2002.08370
  [hep-ph]} \BibitemShut {NoStop}%
\bibitem [{\citenamefont {D\"obrich}\ \emph {et~al.}(2016)\citenamefont
  {D\"obrich}, \citenamefont {Jaeckel}, \citenamefont {Kahlhoefer},
  \citenamefont {Ringwald},\ and\ \citenamefont
  {Schmidt-Hoberg}}]{Dobrich:2015jyk}%
  \BibitemOpen
  \bibfield  {author} {\bibinfo {author} {\bibfnamefont {Babette}\ \bibnamefont
  {D\"obrich}}, \bibinfo {author} {\bibfnamefont {Joerg}\ \bibnamefont
  {Jaeckel}}, \bibinfo {author} {\bibfnamefont {Felix}\ \bibnamefont
  {Kahlhoefer}}, \bibinfo {author} {\bibfnamefont {Andreas}\ \bibnamefont
  {Ringwald}}, \ and\ \bibinfo {author} {\bibfnamefont {Kai}\ \bibnamefont
  {Schmidt-Hoberg}},\ }\bibfield  {title} {\enquote {\bibinfo {title}
  {{ALPtraum: ALP production in proton beam dump experiments}},}\ }\href
  {\doibase 10.1007/JHEP02(2016)018} {\bibfield  {journal} {\bibinfo  {journal}
  {JHEP}\ }\textbf {\bibinfo {volume} {02}},\ \bibinfo {pages} {018} (\bibinfo
  {year} {2016})},\ \Eprint {http://arxiv.org/abs/1512.03069} {arXiv:1512.03069
  [hep-ph]} \BibitemShut {NoStop}%
\bibitem [{\citenamefont {Dolan}\ \emph {et~al.}(2017)\citenamefont {Dolan},
  \citenamefont {Ferber}, \citenamefont {Hearty}, \citenamefont {Kahlhoefer},\
  and\ \citenamefont {Schmidt-Hoberg}}]{Dolan:2017osp}%
  \BibitemOpen
  \bibfield  {author} {\bibinfo {author} {\bibfnamefont {Matthew~J.}\
  \bibnamefont {Dolan}}, \bibinfo {author} {\bibfnamefont {Torben}\
  \bibnamefont {Ferber}}, \bibinfo {author} {\bibfnamefont {Christopher}\
  \bibnamefont {Hearty}}, \bibinfo {author} {\bibfnamefont {Felix}\
  \bibnamefont {Kahlhoefer}}, \ and\ \bibinfo {author} {\bibfnamefont {Kai}\
  \bibnamefont {Schmidt-Hoberg}},\ }\bibfield  {title} {\enquote {\bibinfo
  {title} {{Revised constraints and Belle II sensitivity for visible and
  invisible axion-like particles}},}\ }\href {\doibase 10.1007/JHEP12(2017)094}
  {\bibfield  {journal} {\bibinfo  {journal} {JHEP}\ }\textbf {\bibinfo
  {volume} {12}},\ \bibinfo {pages} {094} (\bibinfo {year} {2017})},\ \bibinfo
  {note} {[Erratum: JHEP 03, 190 (2021)]},\ \Eprint
  {http://arxiv.org/abs/1709.00009} {arXiv:1709.00009 [hep-ph]} \BibitemShut
  {NoStop}%
\bibitem [{\citenamefont {Banerjee}\ \emph {et~al.}(2020)\citenamefont
  {Banerjee} \emph {et~al.}}]{NA64:2020qwq}%
  \BibitemOpen
  \bibfield  {author} {\bibinfo {author} {\bibfnamefont {D.}~\bibnamefont
  {Banerjee}} \emph {et~al.} (\bibinfo {collaboration} {NA64}),\ }\bibfield
  {title} {\enquote {\bibinfo {title} {{Search for Axionlike and Scalar
  Particles with the NA64 Experiment}},}\ }\href {\doibase
  10.1103/PhysRevLett.125.081801} {\bibfield  {journal} {\bibinfo  {journal}
  {Phys. Rev. Lett.}\ }\textbf {\bibinfo {volume} {125}},\ \bibinfo {pages}
  {081801} (\bibinfo {year} {2020})},\ \Eprint
  {http://arxiv.org/abs/2005.02710} {arXiv:2005.02710 [hep-ex]} \BibitemShut
  {NoStop}%
\bibitem [{\citenamefont {Bergsma}\ \emph {et~al.}(1985)\citenamefont {Bergsma}
  \emph {et~al.}}]{CHARM:1985anb}%
  \BibitemOpen
  \bibfield  {author} {\bibinfo {author} {\bibfnamefont {F.}~\bibnamefont
  {Bergsma}} \emph {et~al.} (\bibinfo {collaboration} {CHARM}),\ }\bibfield
  {title} {\enquote {\bibinfo {title} {{Search for Axion Like Particle
  Production in 400-{GeV} Proton - Copper Interactions}},}\ }\href {\doibase
  10.1016/0370-2693(85)90400-9} {\bibfield  {journal} {\bibinfo  {journal}
  {Phys. Lett. B}\ }\textbf {\bibinfo {volume} {157}},\ \bibinfo {pages}
  {458--462} (\bibinfo {year} {1985})}\BibitemShut {NoStop}%
\bibitem [{\citenamefont {Gori}\ \emph {et~al.}(2020)\citenamefont {Gori},
  \citenamefont {Perez},\ and\ \citenamefont {Tobioka}}]{Gori:2020xvq}%
  \BibitemOpen
  \bibfield  {author} {\bibinfo {author} {\bibfnamefont {Stefania}\
  \bibnamefont {Gori}}, \bibinfo {author} {\bibfnamefont {Gilad}\ \bibnamefont
  {Perez}}, \ and\ \bibinfo {author} {\bibfnamefont {Kohsaku}\ \bibnamefont
  {Tobioka}},\ }\bibfield  {title} {\enquote {\bibinfo {title} {{KOTO vs. NA62
  Dark Scalar Searches}},}\ }\href {\doibase 10.1007/JHEP08(2020)110}
  {\bibfield  {journal} {\bibinfo  {journal} {JHEP}\ }\textbf {\bibinfo
  {volume} {08}},\ \bibinfo {pages} {110} (\bibinfo {year} {2020})},\ \Eprint
  {http://arxiv.org/abs/2005.05170} {arXiv:2005.05170 [hep-ph]} \BibitemShut
  {NoStop}%
\bibitem [{\citenamefont {Mariotti}\ \emph {et~al.}(2018)\citenamefont
  {Mariotti}, \citenamefont {Redigolo}, \citenamefont {Sala},\ and\
  \citenamefont {Tobioka}}]{Mariotti:2017vtv}%
  \BibitemOpen
  \bibfield  {author} {\bibinfo {author} {\bibfnamefont {Alberto}\ \bibnamefont
  {Mariotti}}, \bibinfo {author} {\bibfnamefont {Diego}\ \bibnamefont
  {Redigolo}}, \bibinfo {author} {\bibfnamefont {Filippo}\ \bibnamefont
  {Sala}}, \ and\ \bibinfo {author} {\bibfnamefont {Kohsaku}\ \bibnamefont
  {Tobioka}},\ }\bibfield  {title} {\enquote {\bibinfo {title} {{New LHC bound
  on low-mass diphoton resonances}},}\ }\href {\doibase
  10.1016/j.physletb.2018.06.039} {\bibfield  {journal} {\bibinfo  {journal}
  {Phys. Lett. B}\ }\textbf {\bibinfo {volume} {783}},\ \bibinfo {pages}
  {13--18} (\bibinfo {year} {2018})},\ \Eprint
  {http://arxiv.org/abs/1710.01743} {arXiv:1710.01743 [hep-ph]} \BibitemShut
  {NoStop}%
\end{thebibliography}%

\end{document}